\documentclass[aps,twocolumn,amsmath,amssymb]{revtex4}
\usepackage{graphicx}
\usepackage{dcolumn}
\usepackage{bm}
\usepackage{amssymb}
\usepackage{amsmath}

\begin{document}

\title{On the Separability of Attractors in Grandmother Dynamic Systems
with Structured Connectivity}

\author{Luciano da Fontoura Costa}
\email{luciano@if.sc.usp.br}
\affiliation{Instituto de F\'{\i}sica de S\~{a}o Carlos,
Universidade de S\~{a}o Paulo, Av. Trabalhador S\~{a}o Carlense 400,
Caixa Postal 369, CEP 13560-970, S\~{a}o Carlos, S\~ao Paulo, Brazil}

\date{22nd Dec 2006}

\begin{abstract}
The combination of complex networks and dynamic systems research is
poised to yield some of the most interesting theoretic and applied
scientific results along the forthcoming decades.  The present work
addresses a particularly important related aspect, namely the
quantification of how well separated can the attractors be in dynamic
systems underlined by four types of complex networks
(Erd\H{o}s-R\'enyi, Barab\'asi-Albert, Watts-Strogatz and as well as a
geographic model).  Attention is focused on grandmother dynamic
systems, where each state variable (associated to each node) is used
to represent a specific prototype pattern (attractor).  By assuming
that the attractors spread their influence among its neighboring nodes
through a diffusive process, it is possible to overlook the specific
details of specific dynamics and focus attention on the separability
among such attractors.  This property is defined in terms of two
separation indices (one individual to each prototype and the other
considering also the immediate neighborhood) reflecting the balance
and proximity to attractors revealed by the activation of the network
after a diffusive process.  The separation index considering also the
neighborhood was found to be much more informative, while the best
separability was observed for the Watts-Strogatz and the geographic
models.  The effects of the involved parameters on the separability
were investigated by correlation and path analyses.  The obtained
results suggest the special importance of some measurements in
underlying the relationship between topology and dynamics.
\end{abstract}


\maketitle

\vspace{0.5cm}
\emph{`Nothing in excess.' (Delphic proverb)}

\section{Introduction}

Most dynamic systems, from neuronal networks to pattern formation,
involves several interconnected variables, which can be properly
represented in terms of a graph (e.g.~\cite{West01,Bollobas90}) or
complex network (e.g.~\cite{Barabasi:surv, Dorogov:surv, Watts03,
Wasserman94, Newman03, Boccaletti05, Costa:survey}).  Such systems are
henceforth called \emph{complex dynamic systems}.  A state variable
$v(i)$ is normally associated to each node $i$, so that the complete
evolution of the system can be described by the $N
\times 1$ vector $\vec{v}^{(t)}$. In a neuronal network, for instance,
each neuron can be expressed as a vertex (or a node), while synapses
are represented by links (or edges) and the node activation by the
respective state variable.  Once the underlying connectivity of such a
system is represented in terms of its \emph{weight matrix}
$W$~\footnote{$W(j,i)=w$ whenever node $i$ is connected to node $j$ by
an edge with weight $w$.}, several important types of dynamics can be
subsumed as

\begin{equation}  \label{eq_dyn}
\vec{v}^{(t+1)} = f(W \vec{v}^{(t)}).  
\end{equation}

Any layer of the perceptron neuronal network
(e.g. ~\cite{Rosenblatt:1958,Haykin:1999}), for instance, is obtained
by substituting $f()$ by some abrupt function (e.g. hard limit or
sigmoid).  At the same time, the complete classic Hopfield model
(e.g.~\cite{Hopfield:1984,Haykin:1999}) can be represented by this
equation.  A simpler, linear model is obtained by making $f(x)=x$, so
that $\vec{v}^{(t+1)} = W \vec{v}^{(t)}$.  In case $W$ is also a
stochastic matrix (the transition matrix), this linear equation
subsumes all first order Markov chains, a particularly useful and
important dynamic model which is intrinsically associated to random
walks (e.g.~\cite{Rudnik:2004}), Markov chains
(e.g.~\cite{Bremaud:2001}) and diffusion
(e.g.~\cite{Crank:1980,Havlin:2000}).
	
In many dynamic systems (e.g.~\cite{Haykin:1999,Boccara:2004}), the
codification of external stimuli as well as the results from the
network dynamics take place in the $N-$dimensional space defined by
the state variables.  For instance, one particularly pattern may be
represented as the state $\vec{v_1} = [1, 0, -1, 0.5, 1, 0]$, while
another pattern may be associated to $\vec{v_2} = [0, 1, -1, 0, 1,
0]$.  This type of representation is called ~\emph{vector coding}
(e.g.~\cite{Georgopoulos:1986,Salinas:1994}). States which are near
any of such pattern-states are frequently understood to be associated
to that pattern.  For instance, in the case of the example above, a
third pattern similar to $\vec{v_1}$ will tend to be coded as a state
vector $\vec{v_3}$ such that $\delta = ||\vec{v_1}-\vec{v_3}||$ is
small.  Such a smooth, graded coding along the state space is
interesting because it allows for some robustness/redundancy and
flexibility/generality for the representation of the patterns, while
also favoring good dynamic properties such as where the patterns are
accessed through gradient-descent or similar methods
(e.g.~\cite{Haykin:1999,Rudnik:2004}).

However, it is also possible and interesting to consider the situation
where the patterns are associated to specific nodes, not to specific
points in the state space, such as in systems involving the so-called
\emph{grandmother cells} (e.g.~\cite{Barlow:1972, Gross:2002, 
Gazzaniga:1998}).  Extremely important systems including great part of
the mammals cortex are believed to be so organized.  In these cases, a
pattern is associated to a node $i$ such that high values of $v(i)$
signalizes the presence of that pattern.  This type of cells is
ubiquitous in several cortical areas (e.g.~\cite{McIlwain:1996,
Zeki:1990, Zeki:1999}).  For instance, cells which are highly specific
to hands and faces have been found in the inferior temporal
cortex~\cite{Gross:1969, Perrett:1982, Gross:2002}.  It is also known
(e.g.~\cite{Eichenbaum:1989,McIlwain:1996,Zeki:1999,Bosking:2002})
that much of the mammals cortical architecture is characterized by
spatial smoothness, in the sense that cells which are spatially close
tend to exhibit similar dynamics and response (i.e. state
correlations), except at eventual singularities involving
fractures~\cite{Zeki:1999,Bosking:2002}.  Such an organization reminds
of the smooth coding discussed above for dynamic systems where
patterns are represented by the overal state (i.e. vector coding).  It
is important to note that the smoothness property is not exclusive to
topographically organized systems~\footnote{By topographic system it
is henceforth meant that the network nodes have well-defined positions
along an $N-$dimensional space (usually the two-dimensional
space). The terms
\emph{geographic} or \emph{geometric networks} have also been used
in the graph theory and complex networks literatures.} such as the
mammals cortex.  In other words, even in networks where the nodes have
no specific position in an embedding space, neighboring grandmother
nodes may tend to have similar dynamics and response.  In such cases,
it is interesting to use the concept of progressive neighborhoods
around a node (e.g.~\cite{Faloutsos99,Newman_hier01,Cohen03,
Costa04,Costa06,Costa_Silva07})), defined by the immediate neighbors
of a node as well as its second neighbors, and so
on. Following~\cite{Costa04,Costa_Silva07,Costa_Silva07}, in this work
we organize the successive neighborhoods in terms of their
\emph{hierarchical level}.  Therefore, the immediate neighbors of a node $i$
are at hierarchy 1 of $i$, and so on.  In brief, the coding smoothness
property in grandmother systems is reflected by the fact that the
hierarchical neighbors of each node will tend to have responses
similar, though progressively diverging, to that of node $i$.

\begin{figure}[htb]
  \begin{center} 
  \includegraphics[scale=0.5,angle=0]{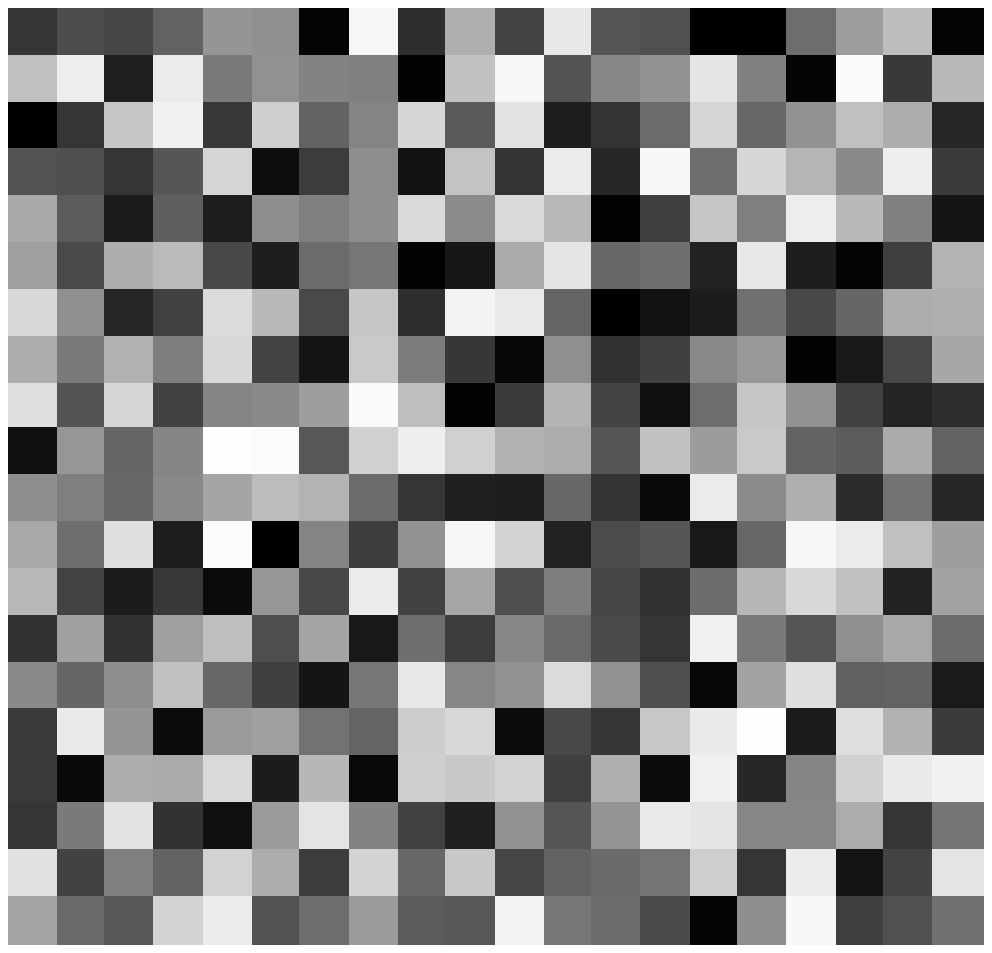} \\
  (a) \\ \vspace{0.3cm}
  \includegraphics[scale=0.5,angle=0]{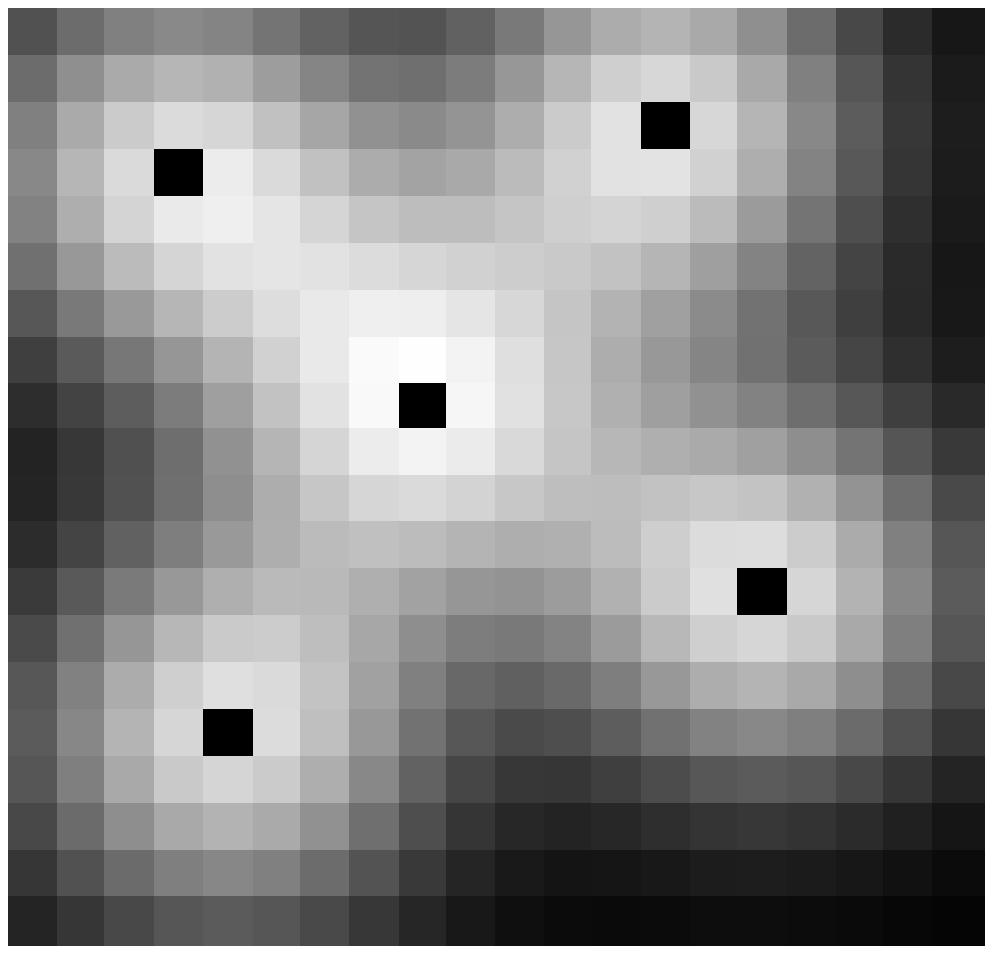} \\
  (b) \\
  \caption{Gray-level visualizations of a topographic state space
  with too many prototypical patterns (a) and of a smooth
  topographic state space (b). Both these spaces are underlined by
  4-neighborhood in the orthogonal lattice. The five prototype
  patterns in (b) have been marked in black for the sake of better
  visualization.~\label{fig:ex}}
  \end{center}
\end{figure}

In dynamic networks with a finite number of nodes, the situation
considered herein, a particularly interesting problem arises as a
consequence of a tension between the grandmother coding and the
smoothness property, a phenomenon directly related to the limited
number of patterns which can be properly represented
(e.g.~\cite{Amit:1993}).  Before proceeding further, it is important
to provide a more objective characterization of such a problem.
Consider that $M$ prototypic, distinct, patterns are to be represented
in the network.  If each of such patterns is represented by a
respective prototypic neuron (node), in the sense that this neuron
will be the most highly activate when that pattern is invoked by the
network, we also need to allow for intermediate cells with graded
replies between pairs of nodes. Observe that such prototype nodes
typically act as attractors for the dynamics of the respective dynamic
system, being accessed, for instance, by using gradient descent and/or
random walks methods. If $M$ is too large, a point will be reached
where each network node will be required to represent each
prototypical pattern, running out of nodes for implementing the graded
responses.  Although this situation occurs at the very limit of the
network capacity, it is important to observe that it may still provide
a complete, invertible representation of the patterns (i.e. an
one-to-one mapping).  On the other hand, the smoothness of the coding
is undermined as the number of prototypical patterns increases.  In
addition to destroying the network potential for generality and
robustness, it will also become impossible to retrieve the patterns by
gradient-descent like mechanisms along the network.
Figure~\ref{fig:ex} illustrates two small topographic state spaces:
one (a) corresponding to the limit situation where each pattern is
associated to each node, the other (b) depicting a smooth map defined
by diffusion around five prototype nodes.  These two spaces were
assumed to have local connectivity, in the sense that each cell
communicates only with its most immediate neighbors in the orthogonal
lattice.  Observe that it is virtually impossible to devise an
effective retrieval or activation mechanism capable of getting to any
specific prototype node in Figure~\ref{fig:ex}(a). In addition, the
failure of any cell will imply in the irrecoverable lost of one
pattern (the origin of the \emph{gradmother cell} concept). On the
other hand, the state space in Figure~\ref{fig:ex}(b) allows a good
level of tolerance to failure (this also means some redundancy), at
the same time as simple dynamic mechanisms such as gradient descent
can be employed for retrieving and activating nodes.

For all that has been said above, it becomes very important to
quantify and study the pattern separability in dynamic systems such as
neuronal networks.  In the case of the dynamic systems which can be
fully represented by Equation~\ref{eq_dyn}, these two properties are
necessarily implied by the network connectivity, defined by the weight
matrix $W$, and the function $f()$.  The present work focuses
precisely on the investigation of the separability between patterns in
grandmother complex dynamic systems.  By considering a diffusive
process emanating from each prototype node, followed by a process
which assigns to each edge the difference between the activation
states at its head and tail nodes, a graded representation of the
prototype patterns can be obtained.  This allows full generality of
the reported investigations as an approximation to many types of
grandmother dynamic systems.  The accessibility to the original
prototypes is simulated through a preferential random walk (diffusion)
starting at every node of the network. After a given relaxation time,
the activation of the network is normalized so as to correspond to
occupancy probabilities and compared to the original prototypes
position.  This is done by considering two separation indices, one
individual, considering the resulting activity only at the prototype
nodes, and the other considering the activation not only at these
nodes, but also at their immediate neighbors.  These two indices, are
henceforth abbreviated as $s_{ind}$ and $s$, are obtained in terms of
the normalized geometric average of the activations at (or around)
each prototype node.  Therefore, the maximum value of these indices
(equal to one) is achieved only when the resulting activation is
totally and uniformly distributed among only the prototype nodes (for
$s_{ind}$) or among these nodes and their immediate neighbros (for
$s$).

The performed simulations and analyses first consider specific network
configurations, in order to gather insights about the influence of each
involved parameter, which are identified and discussed.  Subsequently,
a more systematic investigation of some parameter variations are
performed.  In order to infer the influence of the topologic
features on the dynamics properties (i.e. attractors separation), we
apply correlation analysis and path analysis.  A series of interesting
results are obtained regarding not only the separation indices, but
also the importance of several local and global topologic features
on the overall attractors separation.  Although the current work
concentrates on attractors in grandmother dynamic systems, several of
the results, measurements and methods can be immediately extended to
other relevant systems based on complex networks, including those
involving information transfer and retrieval.

This article starts by presenting a brief review of more closely
related works, and follows by describing the adopted notation and
methodologies.  Next, each of the four considered theoretic network
models are briefly reviewed and their generation described.  The
diffusion procedure and `differentiation' of the state values adopted
to implement the attraction basins are described next, as well as the
diffusion-base mechanism for activation/retrieval of prototypes.  The
two separation indices are motivated and mathematically defined.  The
simulation results and respective discussion are then presented,
followed by the correlation and path analyses of the effects of the
topologic features of the network on the respective attractors
separability.  Several interesting findings and effects are identified
and discussed.  The article concludes with a general synthesis of the
reported investigation and with identification of possible future
developments.

\section{A Brief Review of Previous Developments} \label{sec:review}

The subject of attractor characterization has been extensively
addressed in the literature, so we limit ourselves to reviewing a
small subset of those works which are more general or more directly
related to the investigations being considered in the present work.  

A comprehensive study of attractor networks has been presented by
Amit~\cite{Amit:1992}.  Torres et al. have considered the storage
capacity of attractor neural networks in which the synapses can undergo
depression~\cite{Torres:2002} and found that the memory capacity
decreases with the intensity of the depression. Amit and
Brunel~\cite{Amit:1993} investigated learning in attractor networks
considering the effects of stimuli in a network with wired-up
patterns.  Topographic neuronal networks involving grandmother cells
have been extensively studied by T. Kohonen and collaborators
(e.g.~\cite{Kohonen:1982, Kohonen:2003, Kohonen:2006}) in the form of
the self-organizing map (SOM) or Kohonen networks.  Here, the neurons
are spatially distributed and develop specificity to prototype pattern
stimuli by influencing the weights of neighboring nodes.  Therefore,
spatially adjacent smooth attraction basins are defined.

Several works have brought together the two important areas of complex
networks and dynamic systems, including neuronal networks.  At least
two excellent surveys have been written about this important
issue~\cite{Newman03, Boccaletti05}.  The problem of neuronal
structure and dynamics has also been a subject of growing interest,
including but by no means limited to~\cite{Costa_Salou:2005,
Peichl_Wassle:1983, Costa_BM:2003, SI_BM:2003, GA_hipp:2005,
Ooyen:1995, Ooyen:1999, Ooyen:2001, Ooyen:2004}.  Stauffer et
al.~\cite{Stauffer_etal:2003} investigated the performance of diluted
Hopfield networks underlined by the BA model.  The influence of the
network topology on the dynamics of neural networks has been
investigated also by Costa and Stauffer~\cite{Costa_Stauffer:2003},
who considered spatial neural networks and concluded that its
performance increased with the spatial uniformity of the cells
distribution.  Torres et al.~\cite{Torres:2004}, showed that the
pattern capacity of an attractor neural network with scale free
topology is higher than for a random-diluted network with the same
number of connections.  They also found that, at zero temperature, the
performance of scale free nets improves for larger values of the
power-law exponent. Models of non-randomly diluted neuronal networks
whose connectivity is determined as a function of the shape of its
individual neurons (as well as their relative spatial positions) has
been reported by Costa and
collaborators~\cite{Costa_BM:2003,Costa_EPJB:2004}.  It was found that
the shape of individual neurons can have a great influence on the
respective memory capacity, suggesting that shapes more similar to
real neurons tend to have better attractor properties.  The influence
of the network topology on the recovery of patterns in recurrent
neuronal networks was addressed by Castillo et
al.~\cite{Castillo_etal:2004}. Their work showed that the retrieval
properties can be enhanced by considering connectivity more structured
than in random networks.  Morelli et al.~\cite{Morelli:2004}
investigated the memory capacity in associative networks and found
that the best performance is obtained at an intermediate level of
disorder. Zhou and Lipowski~\cite{Zhou_Lipowski:2005} investigated,
through analytic and simulation means, a general class of dynamic
systems on scale free networks with binary states.  They reported
important variations of performance with respect to the scale free
exponential coefficient. The effect of structured connections on the
interactive statistical mechanics algorithm for minimization of the
Bethe free energy (associated to Ising models) has been studied by
Ohkubo et al.~\cite{Ohkubo_etal:2005}.  The adaptation of the Sznajd
dynamics to take place over the network connectivity instead of its
states has been reported by Costa~\cite{L_sznaj:2005}, yielding
network realizations which are a consequence of an inherent dynamical
process.  Lu et al.~\cite{Lu_etal:2006} considered the effect of
regular, random, small-world and scale free topologies on Hopfield
networks.  They reported, among other findings, that the performance
improved with the local order of the connections, which seems to be in
agreement with~\cite{Costa_Stauffer:2003}. The periodicity of activity
in networks with small-world and scale-free topologies were
investigated by Paula et al.~\cite{Paula_etal:2006} who concluded,
among other findings, that periodic activity appears only for
relatively small networks. Perotti et al.~\cite{Perotti_etal:2006}
studied the interesting problem of associative memory on a growing
diluted Hopfield model which converges to a small-world, scale free
topology and showed that the performance of such a network is higher
than that of a randomly diluted network with the same
connectivity. More recently, Davey et al.~\cite{Davey:2006}
investigated sparse small world associative memory considering
Perceptron training under small world connectivity ranging from local
to global and found that non-symmetric connectivity networks exhibited
superior performance.  By considering random walks as a reference
model for implementing dynamics in complex networks, Costa et
al. investigated the correlations between the topology (node degree)
and activation (frequency of visits to nodes at equilibrium).  They
found that while full correlation is guaranteed for undirected
networks, it can vary substantially in directed networks such as
biological neuronal networks and the Internet, implying that topologic
hubs are not necessarily hubs of activity.  That work also identified
a relationship between scale free networks and the Zipf's
law~\cite{Newman_Zipf:2005}. A study of self-organizing models
underlined by complex networks related to mental processes has been
reported by Wedemann et al.~\cite{Wedemann:2006}. The investigation of
the relationship between topology and dynamics at higher spatial
scales (e.g. cortical areas) has also been addressed in an increasing
number of works (e.g.~\cite{Sporns:2002, Hilgetag:2002,
Hilgetag_etal:2000, Sporns_etal:2000, Sporns_etal:2004,
Costa_Sporns:2006, Costa_Sporns:2007, Costa_etal:2006}).

Increasing attention has also been drawn on the synchronization of
oscillations as the means for pattern encoding and retrieval
(e.g.~\cite{Arecchi:2004}).  Among the works related to the separation
of attractors, Arecchi~\cite{Arecchi:2005} has considered a metric
structure for the percept space, while taking into account the
separation between states.  Wang et al. investigated the influence of
the node degree distribution in the synchronization of two-layer
neural networks.  The criticality of coupling parameters on the
synchronization of an ensemble of identical neural networks with
small-world topology has been addressed by Wang et
al.~\cite{Wang_etal:2005}.

All in all, as far as the influence of connectivity on the performance
of dynamic systems for storing patterns is concerned, several of the
above reviewed works seem to indicate that better results tend to be
obtained by considering non-random connectivity, but at a level of
order that ranges from low to intermediate.

\section{Notation, Basic Concepts and Methodology}

This section covers the concepts and methods used in the current
investigation.  Its subsections present the basic concepts and
measurements in complex networks, the four considered theoretic
network models, the procedure suggested to extend the prototype
influence through their successive neighborhoods in order to establish
the attraction basins, the diffusive way to activate the attractors,
the separation measurements, and the correlation and path analyses
considered in this work.

\subsection{Complex Networks Concepts and Topologic Measurements} 
\label{sec:conc}

A \emph{graph} $\Gamma$ involves a set $V$ of $N$ nodes interconnected
by a set $U$ of $E$ edges, i.e. $\Gamma = (V,U)$.  Each directed edge
linking a node $i$ to a node $j$ is represented as $(i,j)$.  Such a
graph can be conveniently expressed in terms of its \emph{adjacency
matrix} $K$, so that $K(j,i)=1$ whenever an edge $(i,j)$ exists
(otherwise, $K(j,i)=0$).  A graph such that $K(i,j)=K(j,i)=1$ is said
to be non-oriented.  Although the present work focuses on this type of
graph, all reported developments can be extended to directed graphs.
The graphs considered henceforth are also devoid of self-connections
(i.e. $K(i,i)=0$ for every node $i$).  A \emph{complex network} is
henceforth understood as a graph exhibiting a particularly intricated
structure, in the sense of differing from a random graph of the
Erd\H{o}s-R\'enyi type~\cite{ErdosRenyi:1959,ErdosRenyi:1960}.
However, because of statistic fluctuations, a random graph can also
exhibit intricated structure.

Given a generic node $i$, some measurements can be associated to it
(e.g.~\cite{Barabasi:surv,Newman03,Costa:survey}).  Its \emph{degree}
$k(i)$ is defined as the number of edges attached to it, so that

\begin{equation}
  k(i) = \sum_{p=1}^{N} K(p,i). 
\end{equation}

The immediate neighbors of a node $i$ correspond to those nodes which
are directly attached to it. The \emph{clustering coefficient} of a
node $i$ expresses the degree of connectivity among its immediate
neighbors and can be calculated as

\begin{equation}
  cc(i) = \frac{e(i)}{e_{max}},
\end{equation}

where $e(i)$ is the number of undirected edges between the immediate
neighbors of $i$ and $e_{max}$ is the maximum possible number of such
connections, given as $e_{max} = e(i)(e(i)-1)/2$.

The \emph{shortest path} between any two nodes $i$ and $j$ corresponds
to the minimal set $SP$ of adjacent edges connecting those two nodes.
A network in which all nodes can be reached, through paths, from any
other node is henceforth said to be \emph{connected}.  The respective
\emph{shortest path length} $sp$ is defined as the number of edges in $SP$.
Therefore, the immediate neighbors of a node $i$ can be alternatively
defined as the nodes which are at shortest path length of 1 from node
$i$.  The second neighborhood (or neighbors of second hierarchy) are
those nodes which are at shortest path length of 2 from $i$, and so
on.  The maximum shortest path length between any pair of nodes is
defined as the network \emph{diameter}, henceforth abbreviated as
$diam$.

Given a set $R$ of $M$ reference (or seed) nodes, it is possible to
obtain the respective \emph{Voronoi} tessellation~\cite{Stoyan:1996}
of the network~\cite{Costa_Silva07} with respect to the seeds, which
partitions the $N$ nodes into $M$ connected subgraphs associated to
each of the reference nodes.  This can be achieved by, for each node
$i$, identifying which of the reference nodes is closest (in the sense
of shortest path) to $i$ and assigning it to that region.  Given two
nodes $i$ and $j$, whose respective immediate neighbors are
represented by the sets $n(i)$ and $n(j)$, we define their
\emph{common neighbors} as the set $c(i) = n(i) \cap n(j)$.  
Given the Voronoi partition of a network, it is interesting to devise
a measurement capable of expressing the uniformity of the areas $A(i)$
(i.e. number of nodes) of each of the $M$ partitioned regions $i$.
This can be conveniently achieved by considering the \emph{geometric
average} $\left[ A \right]$ of the areas, i.e.

\begin{equation}
  \left[ A \right] = \left( \prod_{i=1}^{M} A(i) \right)^{1/M},
\end{equation}

\begin{figure}[htb]
  \begin{center} 
  \includegraphics[scale=0.55,angle=0]{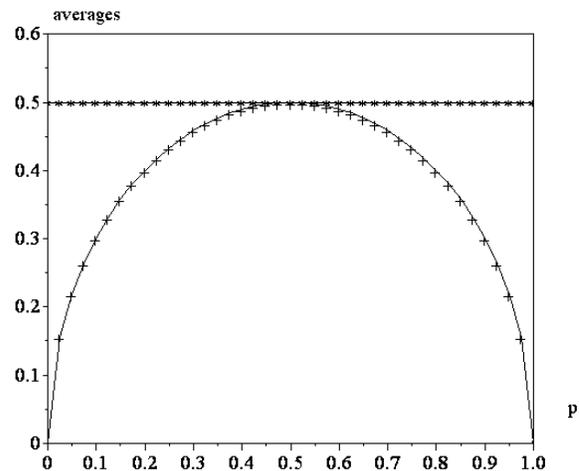} \hspace{0.5cm}
  \caption{The arithmetic ($*$) and geometric ($+$)
  averages between the values $p$ and $q = 1-p$. The
  geometric average allows the quantification of the 
  uniformity between the values of $p$ and $q$, with a
  peak at $p = q = 0.5$.~\label{fig:avgs}}
  \end{center}
\end{figure}

Observe that the geometric average implies a high penalty on higher
variations of the Voronoi areas, therefore providing a more strict
quantification of the homogeneity of those areas.
Figure~\ref{fig:avgs} illustrates the comparison between the
arithmetic and geometric average considering two measurements $p$
(with $0 \leq p \leq 1$) and $q = 1-p$.  The ability of the geometric
average in quantifying the similarity between $p$ and $q$ is evident
from this figure.  While dispersion-related measurements
(e.g. standard deviation and entropy of the state activations) could
be used, they could not be interepreted as quantifications of the
activations.

As the geometric average $u$ for the $M$ Voronoi areas varies as $0
\leq u \leq 1/M$, it is convenient to redefined $\left[ A \right]$ as

\begin{equation}
 \left[ A \right] = M \left( \prod_{i=1}^{P} A(i) \right)^{1/M}.
\end{equation}

The \emph{matching index}~\cite{Sporns:2002,Hilgetag:2002} of a pair
of nodes $i$ and $j$, expressing the relative degree of overlap
between the immediate neighborhoods of those nodes, can be calculated
as

\begin{equation}
  mi(i,j) = \frac{n(i) \cap n(j)}{n(i) \cup n(j)}.
\end{equation}

This measurement gives the fraction of immediate neighbors of $i$ and
$j$ which are common neighbors to them both.  For instance, for the
situation depicted in Figure~\ref{fig:common}, we have that the
matching index of the pair of nodes $P1$ and $P2$ is given as
$m(P1,P2) = 2/8 = 0.25$.

\subsection{Complex Networks Theoretic Models}

Four theoretic models of complex networks are considered in the
present work: Erd\H{o}s-R\'enyi (ER), Barab\'asi-Albert (BA),
Watts-Strogatz (WS) and a geographic model (GG).  Reflecting their
different natures and organizing principles, these models correspond
to a significant portion of the complex networks found in nature,
providing therefore a representative choice of models for the current
study of attractor separability. In order to account for a more
coherent comparison between the separability of attractors in these
four models, each comparison always consider $N$ and $\left< k
\right>$ for each model to be as similar as possible~\footnote{Observe
that it is not possible to ensure identical values of such parameters
in all models. For instance, the average degree $\left< k \right>$
will naturally vary within a limited interval in all models.}.  In
this work, the average degree of the BA model (defined by the
parameter $m$) is always taken as a reference for defining the average
degree of the other models.  The methodology for constructing such
models are described in the following subsections.

\subsubsection{Erd\H{o}s-R\'enyi}

\emph{Random networks}, also called Erd\H{o}s-R\'enyi --
ER, were among the first models of stochastic networks to be
extensively
studied(e.g.~\cite{Rapoport:1957,ErdosRenyi:1959,ErdosRenyi:1960}).
These networks are characterized by constant probability of having a
connection between any of the possible pairs of nodes, and are
therefore related to Poisson processes.  The connectivity of ER
networks can generally be well approximated in terms of its average
degree, implying that such networks are similar to regular networks,
characterized by having the same degree at any of its nodes. As a
consequence of its indiscriminate connectivity and largely regular
organization, the ER type of networks does not typically provide a
good model of natural structures and phenomena, where the connections
tend to follow more purposive and specific rules.

In an ER graph, each possible connection between each possible pair of
nodes, has constant probability $\gamma$ of existence.  Such networks
can be easily created through Monte Carlo simulation by making
$K(i,j)=K(j,i)=1$, with $i \neq j$, with probability $\gamma$.
Observe that the construction of ER networks consider only two
parameters: $N$ and $\gamma$.  In order to obtain values of $\left< k
\right>$ similar to the BA reference, we enforce $\gamma = 2m/(N-1)$,
where $m$ is a parameter of the BA model (see next subsection).

\subsubsection{Barab\'asi-Albert}

The so-called \emph{Barab\'asi-Albert
model}~\cite{Albert:1999,Barabasi:Linked,Barabasi:surv} -- BA, belongs
to the important class of \emph{scale free} networks.  This type of
network is characterized by the fact that the loglog plot of their
node degrees tends to a straight line, implying the absence of any
characteristic scale.  In other words, the node distribution in such a
network follows a \emph{power law}.  One of the most important
properties of scale free networks
(e.g.~\cite{Barabasi:surv,Barabasi:Linked}), when compared to models
such as the ER, is the higher probability of existence of \emph{hubs},
i.e. nodes with particularly high degree.  Such special nodes are
particularly important in defining the connectivity and topologic
features of the network, such as the average shortest path length.
For instance, the fact that a hub connects to many nodes immediately
implies that the shortest path between any of these nodes will be at
most equal to 2 edges.  Several important natural and human-made
structures -- including the Internet, WWW, protein interaction and
even scientific collaborations -- have been found to exhibit the scale
free property (e.g.~\cite{Barabasi:surv,Barabasi:Linked,Newman03}).
The BA model incorporates the so-called
\emph{rich-get-richer} paradigm because of its attachment of links being
preferential to the degree of existing nodes.

In the current work, BA networks are generated starting with $m0$
randomly connected nodes.  At each subsequent step, a new node with
$m$ edges is added to the network, with each of the $m$ edges being
attached to previous network nodes preferentially to their degree.
Therefore, each new connection is more likely to be established with
previous nodes with high degree, implementing the `rich get richer'
paradigm.  As with the ER model, the construction of BA networks also
involves only two parameters: $N$ and $m$.

\subsubsection{Watts-Strogatz}

Historically, the \emph{small world networks} of Watts and
Strogatz~\cite{Watts_Strogatz:1998,Watts03} followed the random
networks of Erd\H{o}s, R\'enyi and collaborators.  Small world
networks are exactly as implied by their name, i.e. the average
shortest path length between their nodes tends to be small.  At the
same time, they also tend to be characterized by relatively high
clustering coefficient, implying that they local connectivity is
relatively high.  The small world property, which has been found to be
present in many interesting networks including ER and BA, has
important implications for the separation of attractors because it
implies that many prototype nodes will be near one another and,
consequently, possibly less separated as far as the dynamics is
concerned.  The WS model considered in this work, however, presents
some specific topologic organization which has potential
implications for the distribution of the prototypes.  More
specifically, this model is characterized by a relatively high
regularity and uniformity of local connectivity.

The Watts-Strogatz networks used in the current work have been
constructed by starting with a ring of nodes where each node is
connected to its $m$ clockwise and anti-clockwise nodes.  After such
an initial network is obtained, $\alpha$\% of the existing connections
are rewired at random.  This network model involves three parameters:
$N$, $m$ and $\alpha$.  All configurations in this work assume
$\alpha = 10$\%.

\subsubsection{A Geographic Model}

Geographic networks (e.g.\cite{Costa_Stauffer:2003, Costa:2003,
Spatial_Satorras:2004, Costa_Diambra:2005, Spatial_Barrat:2005,
Spatial_Newman:2006})) -- also called spatial, geometric or topologic
-- are characterized by the fact that their nodes have well-defined
spatial positions within an embedding space.  Frequently, the
connectivity in such networks is considered to be highly influenced by
the spatial adjacencies and/or spatial proximity between its nodes, in
the sense that two nodes which are adjacent or near one another will
have higher chances of being connected.  Therefore, geographic models
in small dimensional spaces (e.g. 2D or 3D) tend not to be small
world.  Such a property is particularly important as far as the
separation of the prototype nodes is concerned because, in principle,
this type of network allows more space distribution of nodes which are
relatively further apart.  Here, we consider one of the simplest
possible approaches to obtaining a geographic model, which involves
the distribution of the $N$ nodes uniformly along a 2D space followed
by the interconnection of all nodes which are closer than a given
distance $d$.

In order to build a GG network, we start with an empty $L \times L$
discrete space $S$, such that each of its positions is expressed as
$S(x,y)$, where $i$ and $j$ are integer values so that $1 \leq x, y
\leq N$.  This space is henceforth understood as a Poisson 
field with density $\rho$, in the sense that any region with area $a$
(i.e. number of discrete elements $(x,y)$) will have, in the average,
a total of $a \rho$ points marked as $S(x,y)=1$~\cite{Stoyan:1996}.
The network nodes are selected by considering each position $(x,y)$ in
the space $S$ with probability $\rho = N/L^2$, implying an average
total number of nodes $N$.  Then, each of such nodes, marked as
$S(x,y)=1$, is connected to all other nodes to be found up to a
maximum Euclidean distance $d$.  Therefore, the average degree of the
network can be defined by controlling $d$.  More specifically, we make
$d = L\sqrt{2m/(N \pi)}$, so that every disk of radius $r$ in $S$
centered at each node will contain, in the average, $\left< k \right>
= 2m$, where $m$ is the BA parameter taken as the reference.  The
growth parameters of such a network model therefore are again limited
to only $N$ and $m$.  In order to ensure a relatively small variation
of $N$, every generated network with a total number of nodes smaller
than 90\% of the desired value of $N$ were discarded.
Figure~\ref{fig:exs} illustrates a GG complex network obtained by
using the methodology described above for $N=50$ and $m=3$.

\subsection{Generating the Attraction Basins}

As indicated in the Introduction, in order to avoid the intricacies
and specificities of how each distinct dynamic system represents
attractors, here we resort to a simple methodology involving diffusion
of activity from the prototype nodes, followed by the transformation
of the so obtained activity into a derivative
network~\cite{Costa_simple:2004,Costa_Travieso:2005}.
Although not reproducing in detail the attraction basins which would
be otherwise produced by diverse specific dynamics, this approach does
ensure the smoothness of coding, i.e. the property that nodes which
are topologic close will tend to have similar state dynamics.  The
details of such a methodology are presented as follows.

Let the $M$ prototype patterns be associated with respective nodes
chosen with uniform probability among the $N$ network nodes.  Let
$\vec{P}$ be the vector such that $P(i)=1$ if and only $i$ is one of
the prototype nodes, with $P(i)=0$ otherwise.  In order to allow a
probabilistic interpretation, we normalize this vector as $\vec{p} =
\vec{P} / \sum_{i=1}^{N} p(i)$.  This vector will act as the fixed
source of probabilities during the diffusion.  The adjacency matrix
describing the network is also normalized into its respective transfer
matrix $W$, i.e.:

\begin{equation}
  W(i,j) = K(i,j) / k(i), 
\end{equation}

where $k(i)$ is the degree of node $i$. Note that all the sums of $W$
along each of its columns will now be equal to 1, i.e. $W$ is a
\emph{stochastic matrix}.  In order to ensure that the matrix $K$ is
connected (irreducible)~\cite{Bremaud:2001}, all the nodes which do
not belong to the main connected component are excluded from the
network at the end of its respective construction.  Because the
considered average node degrees are relatively high, and well above
the percolation critic density, very few nodes are removed through
such a procedure.

Now, the final distribution of occupancy $\vec{\Omega^{T}}$ of each
node $i$ after $T$ interactions can be calculated by applying
recursively ($T$ times) the following set of equations:

\begin{eqnarray} 
  \vec{a} = \left( \vec{\Omega}^{t} + \vec{p}  \right),    \nonumber \\
  \vec{b} = \left( \vec{a} / \sum_{i=1}^{N} a(i) \right),   \nonumber \\
  \vec{\Omega}^{t+1} = W \vec{b}.
\end{eqnarray}

The number $T_t$ of total interactions is henceforth defined as
corresponding to $3$ times the diameter of the respective network,
i.e. $T_t = 3 diam$.  Figure~\ref{fig:exs}(b) illustrates the
occupancy states obtained for a GG network with $N=100$, $\left< k
\right> = 6$ and $M=2$ after $T_t=27$ interactions.

After the occupancy state of each node is obtained for the network,
its respective \emph{derivative network} $\Delta$
(e.g.~\cite{Costa_simple:2004, Costa_Travieso:2005}) is obtained by
applying the equation below for each edge $(i,j)$ existing in the
original network.

\begin{eqnarray}
  aux = \Omega(j) - \Omega(i)  \nonumber \\
  \left\{ \begin{array}{ll}
    aux>0  \Rightarrow &  \Delta(j,i) = aux  \nonumber \\
    aux<0  \Rightarrow &  \Delta(j,i) = aux/10  \nonumber \\
    aux=0  \Rightarrow &  \Delta(j,i) = -1  \nonumber \\
    \end{array}
  \right.
\end{eqnarray}

The following substitution is applied afterwards:

\begin{equation}  \label{eq:subst}
    \Delta(j,i) = -1  \Rightarrow  \Delta(j,i)=max(\Delta)/\beta.
\end{equation}

All cases in this work assumes $\beta=1000$, but this parameters has
been found not to be critical.  

Note that the values $\Delta(j,i)$ correspond to the weights of the
respective edges $(i,j)$.  Those edges which connect a node with small
induced activity to a node with higher activity will have larger
weights.  More specifically, in case $aux>0$ the weights are directly
proportional to the difference of activations.  Edges leading from
higher to smaller activations have one tenth of the reciprocal edge.
Because several adjacent nodes may result with equal activations, the
eventual previous connectivity between then will be preserved though
at an incremental value proportional to the maximum activation
(Equation~\ref{eq:subst}).  Observe that the matrix $\Delta$ is not
symmetric.

\subsection{Activating the Network}

Once the derivative network defined by the weight matrix $\Delta$ has
been calculated, its activation can be easily achieved through a
random walk preferential to the weights of the edges (see
also~\cite{Toro:2004}, where information is transmitted considering
the gradient of a node).  In order to do so, we obtain the stochastic
version of $\Delta$ by applying the following equation to each of its
edge $(i,j)$

\begin{equation}
  \delta(j,i) = \Delta(j,i) / \sum_{j=1}^{N} \Delta(j,i).
\end{equation}

\begin{figure*}[htb]
  \begin{center} \includegraphics[scale=0.75,angle=0]{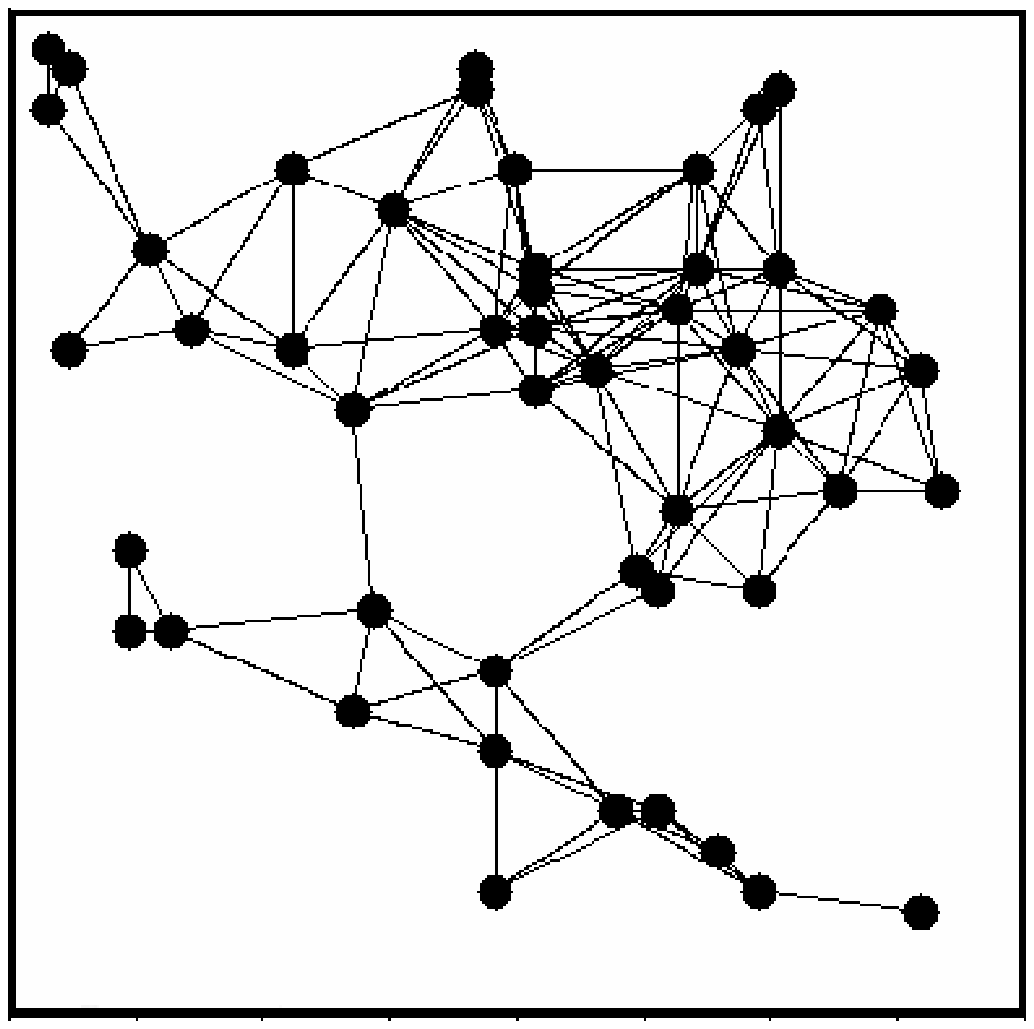}
  \hspace{0.5cm} \includegraphics[scale=0.75,angle=0]{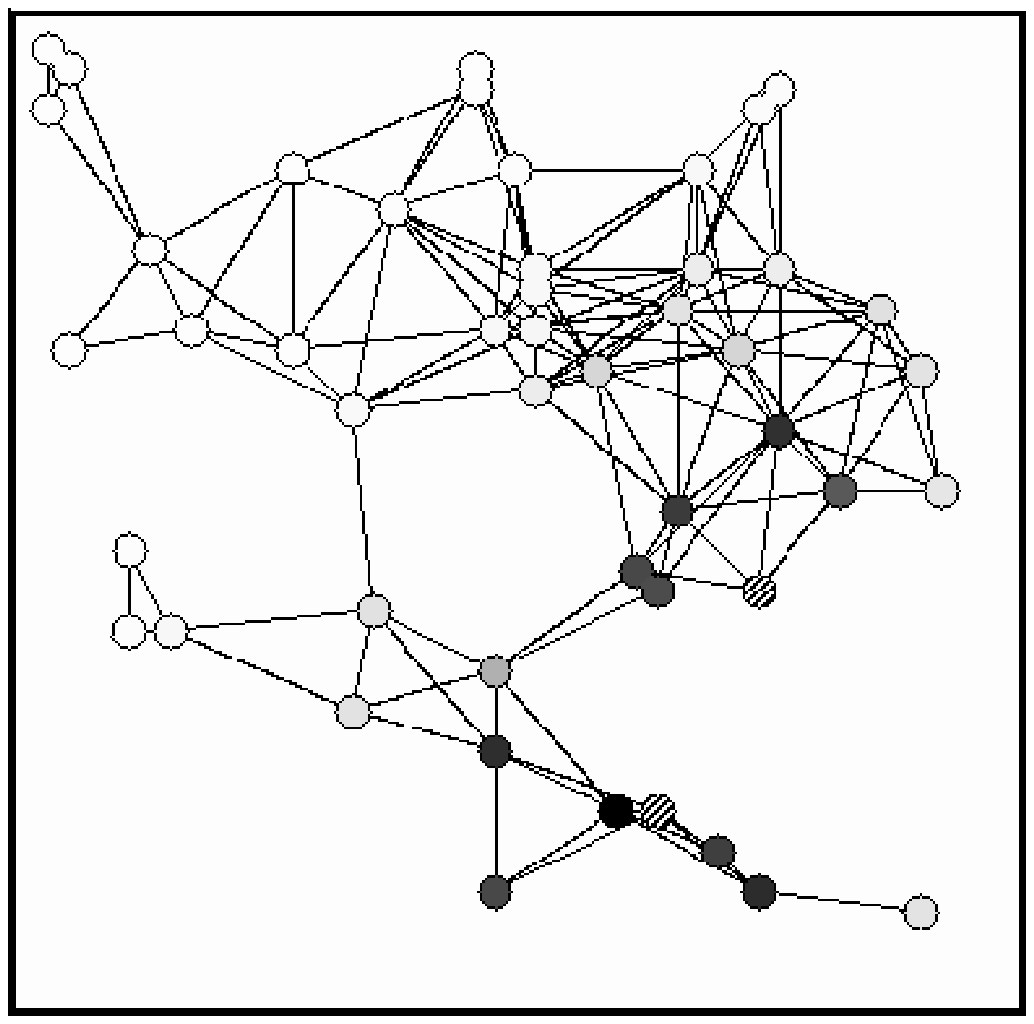} \\
  (a) \hspace{8cm} (b) \vspace{0.5cm} \\
  \includegraphics[scale=0.75,angle=0]{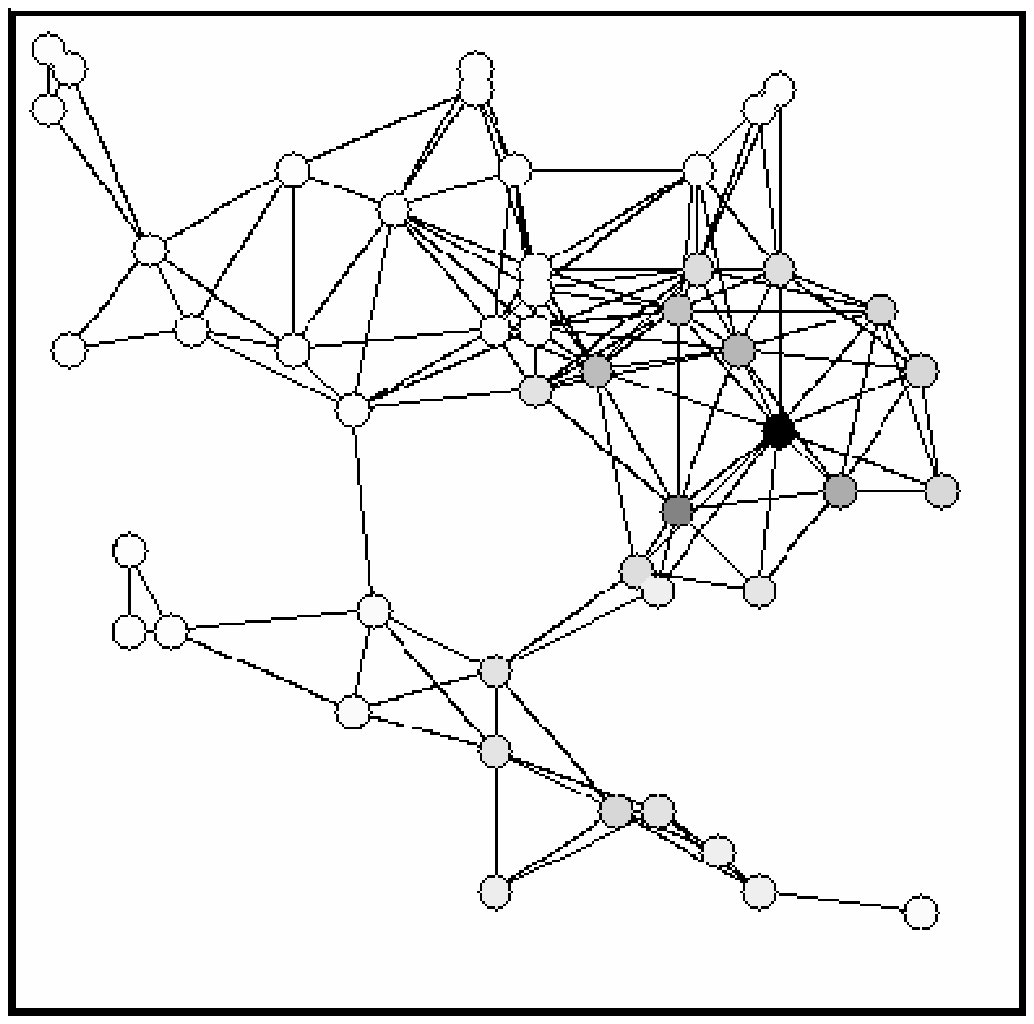} \hspace{0.5cm}
  \includegraphics[scale=0.75,angle=0]{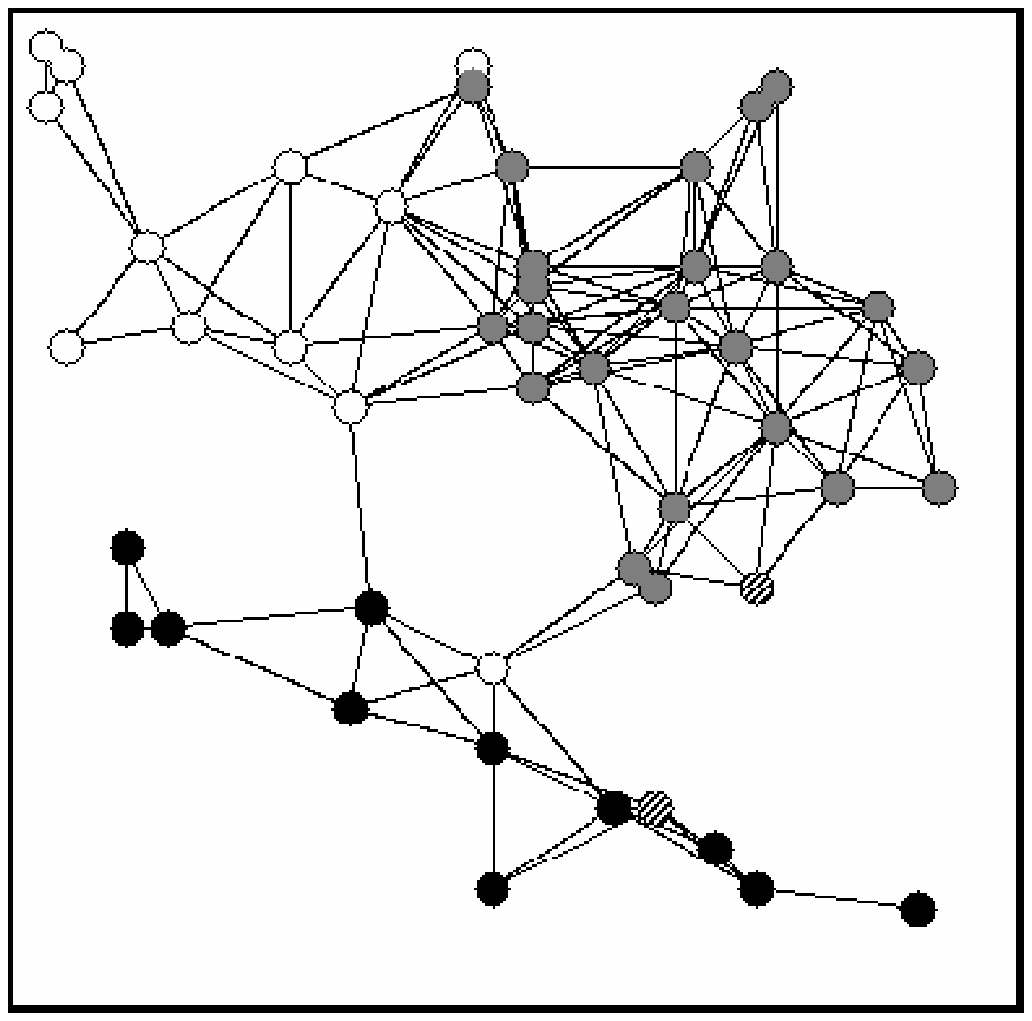} \\ (c) \hspace{8cm}
  (d) \\ \caption{Example of GG network obtained for $N=50$ and $m=3$
  (with $\left< k \right> = 5.5$) (a).  The
  respective occupancy states, defining the basis of the prototype
  nodes (striped), after $27$ interactions (b). Note the smooth
  distribution of states through both the topologic neighborhoods. 
  The activation induced by the random walks
  preferential to the weights of the respective derivative network
  (c).  The Voronoi tessellation defined by the two prototype nodes
  (represented as striped) (d).~\label{fig:exs}} \end{center}
\end{figure*}

The activation after $T$ steps is now given as

\begin{equation}
  \vec{\alpha} = 1/N \delta^T \vec{1},
\end{equation}

where $\vec{1}$ is the $N \times 1$ vector of ones.  As before, we
assume that the total number of interactions $T_r$ is equal to 3 times
the network diameter.  Therefore, the final activation corresponds to
the near equilibrium occupancy of each node after starting from any
node.  Figure~\ref{fig:exs}(c) shows the activations obtained by the
method described above with respect to the network in
Figure~\ref{fig:exs}(b).  It is interesting to note that the peaks of
the obtained activations do not necessarily correspond to the original
prototype nodes.  This is an interesting consequence of the fact that
the activation of the derivative network is affected not only by the
attraction basins, but also by the
\emph{in-strength}~\footnote{In a directed weighted network such as the 
derivative network, the \emph{in-strength} of a node $i$ is defined as
being equal to the sum of the weights of each inbound edge.} and the
local connectivity of each node.  More
specifically~\cite{Costa_etal:2006}, in case the in-strength is equal
to the out-strength for all nodes, the equilibrium activation will be
directly proportional to the respective in-strength.  However, this
result is not guaranteed for asymmetric connectivity such as in the
derivative network.

\subsection{Separation Indices}

Given a specific grandmother dynamic system, it is important to
quantify in an objective manner the separability of its attractors.
Recall that the system incorporates $M$ prototype patterns, associated
to respective prototype nodes $p = 1, 2,
\ldots, M$.  In this work we propose two separation indices,  
$s_{ind}$ and $s_{ngh}$, defined by taking into account diffusion
dynamics along the complex network underlying the dynamic system.  For
simplicity's sake, the latter is henceforth called simply as
\emph{separation index} and abbreviated as $s$.  

Once the activations have been obtained in each of the dynamic states
associated to the nodes, we define the \emph{individual separation
index} $s_{ind}$ as being proportional to the geometric average of the
activations $v(p)$ at each prototype node $p = 1, 2,\ldots, M$, i.e.

\begin{equation}
  s_{ind} = M \left( \prod_{i=1}^{M} v(i) \right)^{1/M},
\end{equation}

where the proportionality factor $M$ is introduced in order to ensure
that $0 \leq w \leq 1$ instead of $0 \leq w \leq 1/M$.  As with the
quantification of the uniformity of the Voronoi areas
(Section~\ref{sec:conc}), the geometric average will reach its maximum
when all activations $v(i)$ have the same value.

\begin{figure}[htb]
  \begin{center} \includegraphics[scale=0.5,angle=0]{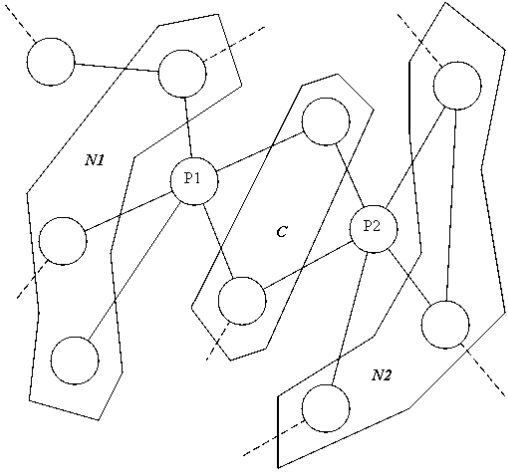}
  \caption{A portion of a hypothetical network showing two prototype
  nodes ($P1$ and $P2$).  The nodes belonging to both immediate
  neighborhoods are identified by the region $C$, while the nodes
  belonging to the respective immediate neighborhoods minus the common
  nodes are enclosed by the regions $N1$ and $N2$,
  respectively.~\label{fig:common}} \end{center}
\end{figure}

However, as preliminary simulations (see Section~\ref{sec:fixed})
showed that this index tends to be too small, an alternative separation
index has been considered which takes into account also the immediate
neighborhoods of each prototype node.  Consider the situation
illustrated in Figure~\ref{fig:common}.  This figure shows two
prototype nodes $P1$ and $P2$ as well as their common $C$ immediate
neighboring nodes.  Because these nodes are at the same shortest path
distance from $P1$ and $P2$, they do not contribute to the
discrimination between those prototype nodes and are therefore not
considered in the calculation of the neighborhood separation index,
which is more formally defined as follows.

Let $Z(i)$ be the set including the respective prototype node $i$ and
its immediated neighbors which are not common to the immediated
neighborhoods of any of the other prototype nodes.  The probability
$p(i)$ at this set of nodes corresponds to the sum of the normalized
activations of the states in $Z(i)$.  Ideally, all prototype nodes
should result with probability equal to $1/M$, implying that all
prototypes are equally accessible and therefore maximally separable.
In order to quantify how the obtained network state approaches such a
reference separation, we define the neighborhood separation index
$s_{ngh}$ as corresponding to the geometric average of the total
probabilities at each set $Z(i)$, for all $M$ prototype nodes $i$,
i.e.

\begin{equation}
  s = s_{ngh} = M \left( \prod_{i=1}^{M} p(i) \right)^{1/M},
\end{equation}

where the multiplying constant $M$ is, as before, included in order to
imply that $0 \leq s \leq 1$.  The maximum separation between
attractors is therefore obtained whenever $s=1$.

\subsection{Correlation Analysis}

Given several measurements of the topology of the network under
analysis, as well as the separation indices, a first insight about
their possible relationship and redundancy can be obtained by
considering the \emph{Pearson correlation coefficient}.  Let us
express each measurement as a random variable $X$ of which we have
$N_X$ observations.  First, we
\emph{standardize}~\cite{Johnson_Wichern:2002} the variable $X$ as

\begin{equation}
  \tilde{X} = \frac{X - \mu_X}{\sigma(X)},
\end{equation}

where $\mu_X$ and $\sigma(X)$ are the estimate of the average and
standard deviation of $X$.  The new, normalized variable has average
zero and unit variance. The Pearson correlation coefficient $r(X,Y)$
between any pair of normalized measurements $X$ and $Y$ can now be
estimated as

\begin{equation}
  r(X,Y) = \frac{1}{N_X-1} \sum_{i=1}^{N} \tilde{X} \tilde{Y}.
\end{equation}

Observe that $-1 \leq r(X,Y) \leq 1$, while $r(X,Y)=0$ means lack of
correlation between the two measurements.  It is important to stress
that uncorrelation does not imply statistic independence.  At the same
time, correlation can by no means be understood as a certain
indication of \emph{causality}.  In case two measurements result with
high absolute value of the Pearson correlation coefficient, they are
said to be \emph{correlated}, indicating redundancy of measurements.
However, even correlated measurements can contribute to the
characterization and discrimination between the
networks~\cite{Costa:survey}.

\subsection{Path Analysis}

While the Pearson correlation coefficient quantifies, in a normalized
fashion, the joint variation of two measurements, such a pairwise
measurement does not consider information about additional
measurements.  A series of sound statistic methods, ranging from
\emph{multivariate regression} (e.g.~\cite{Kleinbaum:1998,
Montgomery:2001}) to the more sophisticated
\emph{Structural Equation Modeling} (SEM) (e.g.~\cite{Raykov:2000, 
Kline:2005}), can be considered in order to obtain a more
representative characterization of the relationship between multiple
random variables (including latent variables) or measurements.  In
this work we considered the \emph{Path Analysis} methodology
(e.g.~\cite{Wright:1920, Raykov:2000, Kline:2005}), understood here as
a particular case of the SEM framework, in order to obtain indication
about the influence of the several measurements of the topology of the
networks on the respective attractor separation indices.

\begin{figure}[htb]
  \begin{center} 
  \includegraphics[scale=0.7,angle=0]{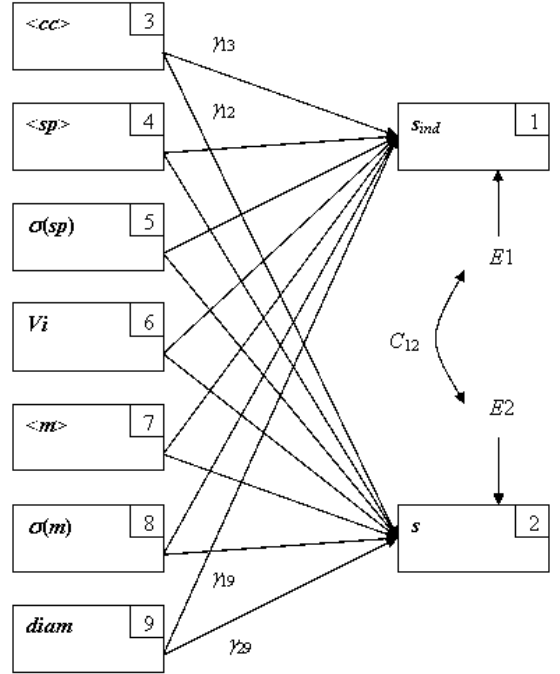} 
  \caption{The structural relationship between the topologic
  and dynamic measurements considered for the path analysis
  reported in this work.  The covariances between variables
  3 to 9 are not shown for simplicity's sake.~\label{fig:diagram}}
  \end{center}
\end{figure}

Path analysis was largely developed by S. Wright
(e.g.~\cite{Wright:1920}) in order to model explanatory relationships
between observed variables.  This methodology is similar to the
solution of a system of equations implied by substituting the model
generated covariance matrix into the sample
covariance~\cite{Raykov:2000}.  One of the interesting features of
this approach is that it considers the influence of the covariances
between all variables, sometimes being closely related to multivariate
linear regression.  The path analysis performed in this work considers
the structural relationship between the topologic and dynamic
properties of the investigated networks as shown in
Figure~\ref{fig:diagram}.  A relatively simple relationship between
the measurements has been considered, where the topologic features are
understood to cause the dynamic properties, namely the individual and
neighborhood attractor separation indices ($s_{ind}$ and $s$,
respectively).  Each of the measurements are associated to a reference
number (upper righthand side of each box).  The parameters
$\gamma_ij$, which are in principle not known, express the importance
of the topologic measurements with respect to the separation indices.

Equations~\ref{eq:sind} and~\ref{eq:s} express the relationship
between the considered variables, reflecting the fact that the
regression coefficients $\gamma_ij$ establish the weights of the
influences of each topologic variable onto the two separation
indices. The environment
LISREL~\footnote{http://www.ssicentral.com/lisrel/index.html} was used
in this work in order to perform path analysis.

\begin{eqnarray} 
  s_{ind} \propto \gamma_{13}\left<cc\right> + \gamma_{14}\left<sp\right>
  +\gamma_{15}\sigma{sp} + \gamma_{16}Vi \nonumber \\
  +\gamma_{17}\left<mi\right> + \gamma_{18}\sigma(mi) +\gamma_{19}diam  
  \label{eq:sind} \\ 
  \nonumber \\ s \propto \gamma_{13}\left<cc\right> +
  \gamma_{14}\left<sp\right> +\gamma_{15}\sigma{sp} +
  \gamma_{16}Vi \nonumber \\ +\gamma_{17}\left<mi\right> +
  \gamma_{18}\sigma(mi) +\gamma_{19}diam   \label{eq:s}
\end{eqnarray}

\section{Results and Discussion}

This section starts by identifying each involved parameter which can
affect the simulations and follows by presenting the results obtained
for specific configurations of $N$ and $m$ and an analysis of the
variation of the parameters.  The statistic analysis of the
relationship between the topologic and dynamic measurements by using
correlation and path analyses is also reported.

\subsection{Involved Parameters and Their Expected Effects}

An important first issue to be considered while investigating the
attractors separation in complex dynamic networks concerns the
identification of all involved parameters which can influence the
results.  The parameters involved in our simulations, as well as their
expected effects, are described in the following:

\emph{The network model and its intrinsic parameters:} Each theoretic
model of complex network considered in this work is likely to yield
different attractors separations.  Those networks characterized by the
small world property (i.e. ER, BA and WS) are, in principle, likely to
produce less separated attraction basis because of the relatively
small shortest path distance, expected in the average, between the
prototype nodes (however, see Section~\ref{sec:res_path})).  While the
ER, BA and GG models can be generated by considering just two
parameters (i.e. $N$ and $m$), the WS network also involves a third
parameters corresponding to the percentage of rewirings, which is
fixed as $\alpha = 10$\%.

\emph{The network size $N$:} This is an important global parameter of
every complex network, corresponding to its total number of nodes.
One of the most important aspects related to this parameter are the
so-called \emph{finite size effects}, namely the fact that the network
properties change considerably when moving from large (possibly
infinite) to small values of $N$.  As an extreme example, the
connectivity of any network will decrease when $N$ approaches just one
or two nodes.  Although in this work we are more interested in finite
size networks, it is often useful to try to extrapolate from
properties measured for smaller values of $N$ to the infinite limit,
as done in Section~\ref{sec:effs}.  Because of the computational
demand required to simulate hundreds of realizations for each
configuration, the current work is limited to relatively small values
of $N$.  As far as the attractors separation is concerned, it is
expected that, for fixed $m$ and $M$, the separability will tend to
increase with $N$ because of the additional space thus allowed for the
representation and distribution of the prototypes.

\emph{The network average node degree $\left< k \right>$:}  
This parameter, which in this work is defined with respect to the
reference $m$, is particularly important in defining the overall
degree of connectivity in the network, especially in those structures
which are not scale free and therefore have specific degree scales.
Generally, smaller values of $\left< k \right>$ tend to imply longer
shortest paths, possibly improving the attractors separation (however,
see Section~\ref{sec:res_path}).  This parameter is not systematically
investigated in this work, which mostly considers fixed $m = 3$
(however, a situation with $m=10$ is considered in
Section~\ref{sec:fixed}).

\emph{The number $M$ of prototype patterns to be represented:} 
This parameter is directly involved in the separability of the
prototype patterns, in the sense that the larger the number of
prototypes, the less separated they tend to be.  

\emph{The number $T_t$ of interactions used for diffusion around each
prototype node:} The diffusion of activity emanating from the
prototype nodes has been verified to converge quickly for all
considered networks, so that the assumption of the total number of
interactions to be given as $T_t = 3 diam$ practically ensures the
resulting activity to correspond to its equilibrium state.

\emph{The number $T_r$ of interactions considered in the retrieval
random walk:} Again, the assumed total number of interactions used in
the activation of the attraction basins is large enough to imply near
equilibrium states.

\subsection{Fixed Configurations} \label{sec:fixed}

In order to get a better understanding about the separability of the
attractors in the considered four theoretic models and to obtain
preliminary indications about the effects of the involved parameters,
we considered a set of preliminary simulations as described in the
following.  All results presented in this section considered
$T_t=T_r=3 diam$ and 500 realizations of each configuration.

\begin{figure*}[h]
  \begin{center} 
  \includegraphics[scale=0.55,angle=0]{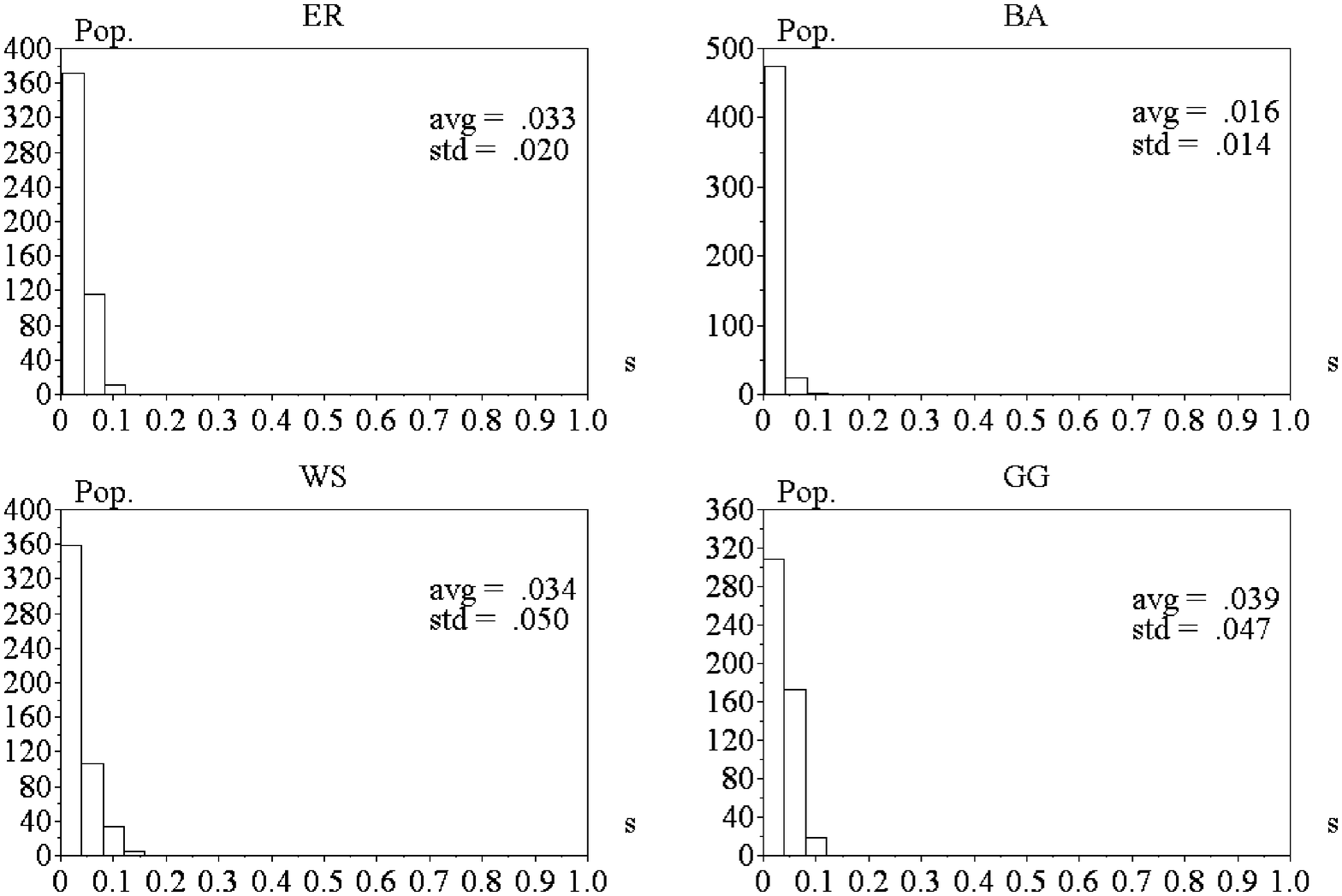}
  \caption{The population histograms of the separation index $s_{ind}$
  for the four considered theoretic models of complex networks with
  $N=100$, $M=3$ and $\left< k \right> = 6$ (i.e. $m=3$).  It is clear
  from these histograms that the separation index $s_{ind}$ does not
  provide good resolution.~\label{fig:caso_sind}}
\end{center}
\end{figure*}

\begin{figure*}[htb]
  \begin{center} 
  \includegraphics[scale=0.55,angle=0]{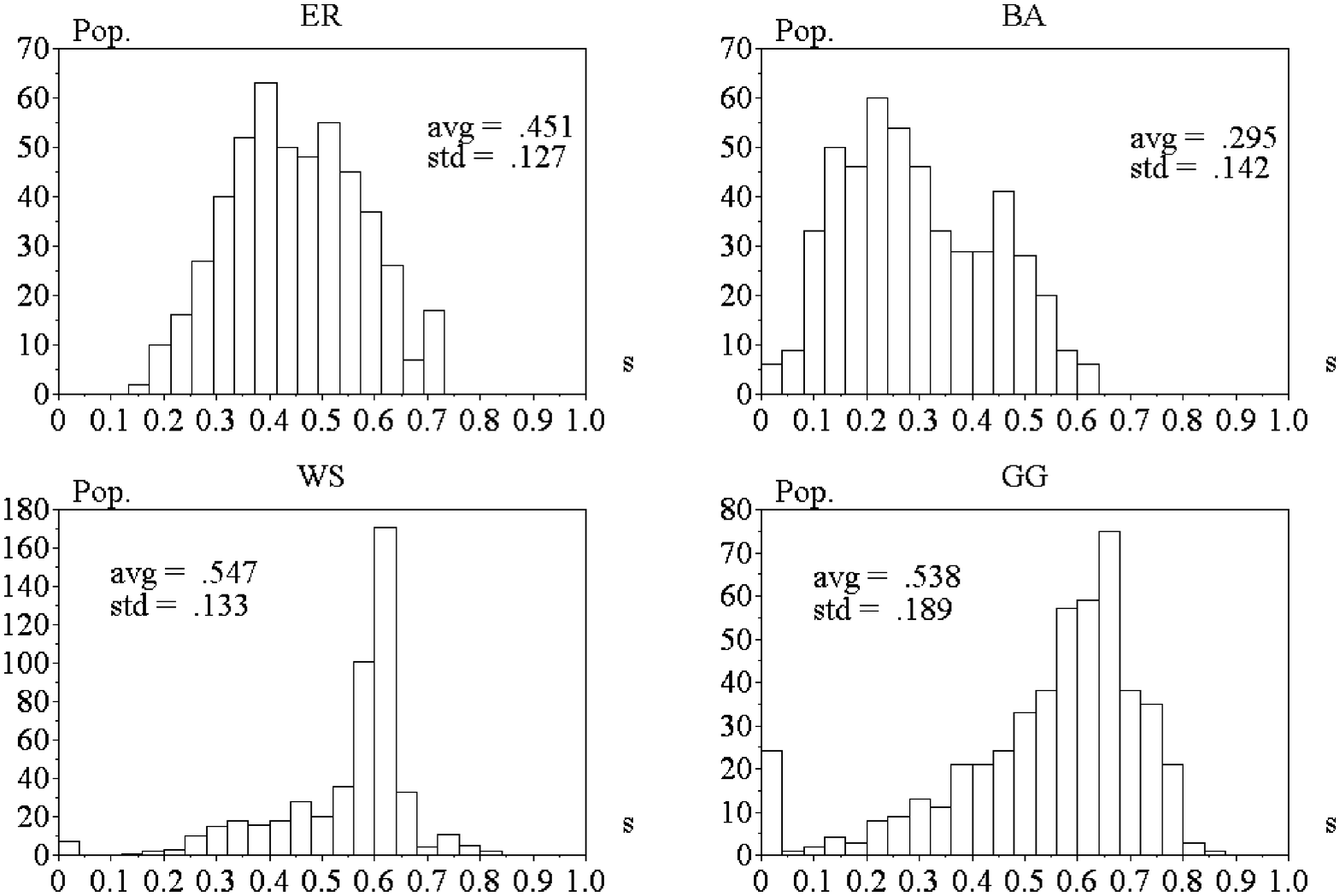} 
  \caption{The population histograms of the separation index $s$
  for the four considered theoretic models of complex networks
  with $N=100$, $M=3$ and $\left< k \right> = 6$.
  Similar results were obtained for the ER/BA and WS/GG models.
  The former case involves smaller average and standard deviation of
  the $s$ index.~\label{fig:caso1}}
  \end{center}
\end{figure*}

So as to have the first glimpses of the separability in each of the
four complex networks models, we considered $N=100$, $\left< k \right>
= 3$, and a relatively small number of prototypes $M=3$.  The
individual separation indices $s_{ind}$ and $s$ obtained for each of
the four models are shown in terms of their respective population
histograms in Figures~\ref{fig:caso_sind} and~\ref{fig:caso1}.  The
respective averages and standard deviations are also shown inside each
graph.  The first important result is that the individual separation
index $s_{ind}$ allows considerably less resolution than the
neighborhood index $s$, with most of its values being close to zero
(Fig.~\ref{fig:caso_sind}).  As this effect has been confirmed for
other configurations, we henceforth focus attention on the
neighborhood separation index $s$.  Interestingly, similar
distributions have been obtained for the pairs of models ER/BA and
WS/GG in Figure~\ref{fig:caso1}, the former being characterized by
smaller averages of $s$.  Observe also the presence of cases where
$s=0$ in the WS and GG cases.  The largest average, implying the
better average attractor separability, was obtained for the WS model,
followed by the GG networks, which also implied a significant number
of null separability indices.  This phenomenon is caused whenever all
immediate neighbors of the prototype nodes are common.  These results
also show that a reasonable separation between the three prototype
attractors have been obtained for the WS and GG cases.

\begin{figure*}[htb]
  \begin{center} 
  \includegraphics[scale=0.55,angle=0]{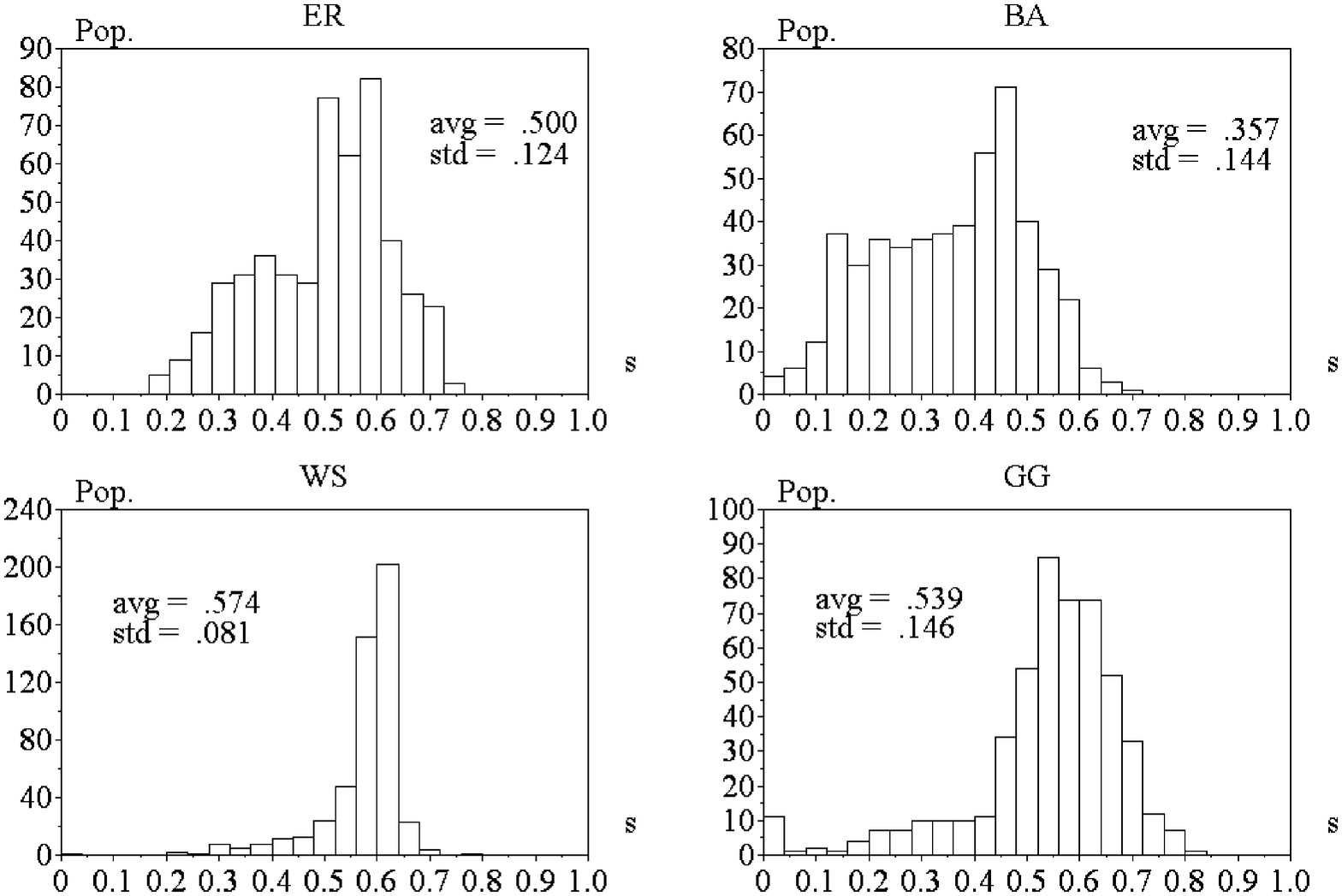} 
  \caption{The population histograms of the separation index $s$
  for the four considered theoretic models of complex networks
  with $N=200$, $M=3$ and $\left< k \right> = 6$.
  Similar results were obtained for the ER/BA and WS/GG models.
  The results are similar to those obtained for $N=100$ (see
  Figure~\ref{fig:caso1}).~\label{fig:caso2}}
  \end{center}
\end{figure*}

Next, we verify the possible influence of the network size $N$ by
considering the same configuration as before, but with $N=200$.  The
respective results are shown in Figure~\ref{fig:caso2}.  It is clear
from these results that the larger size of the network had relatively
little influence on the separation indices, suggesting the finite size
effect to be small for this number of parameters.  

\begin{figure*}[htb]
  \begin{center} 
  \includegraphics[scale=0.55,angle=0]{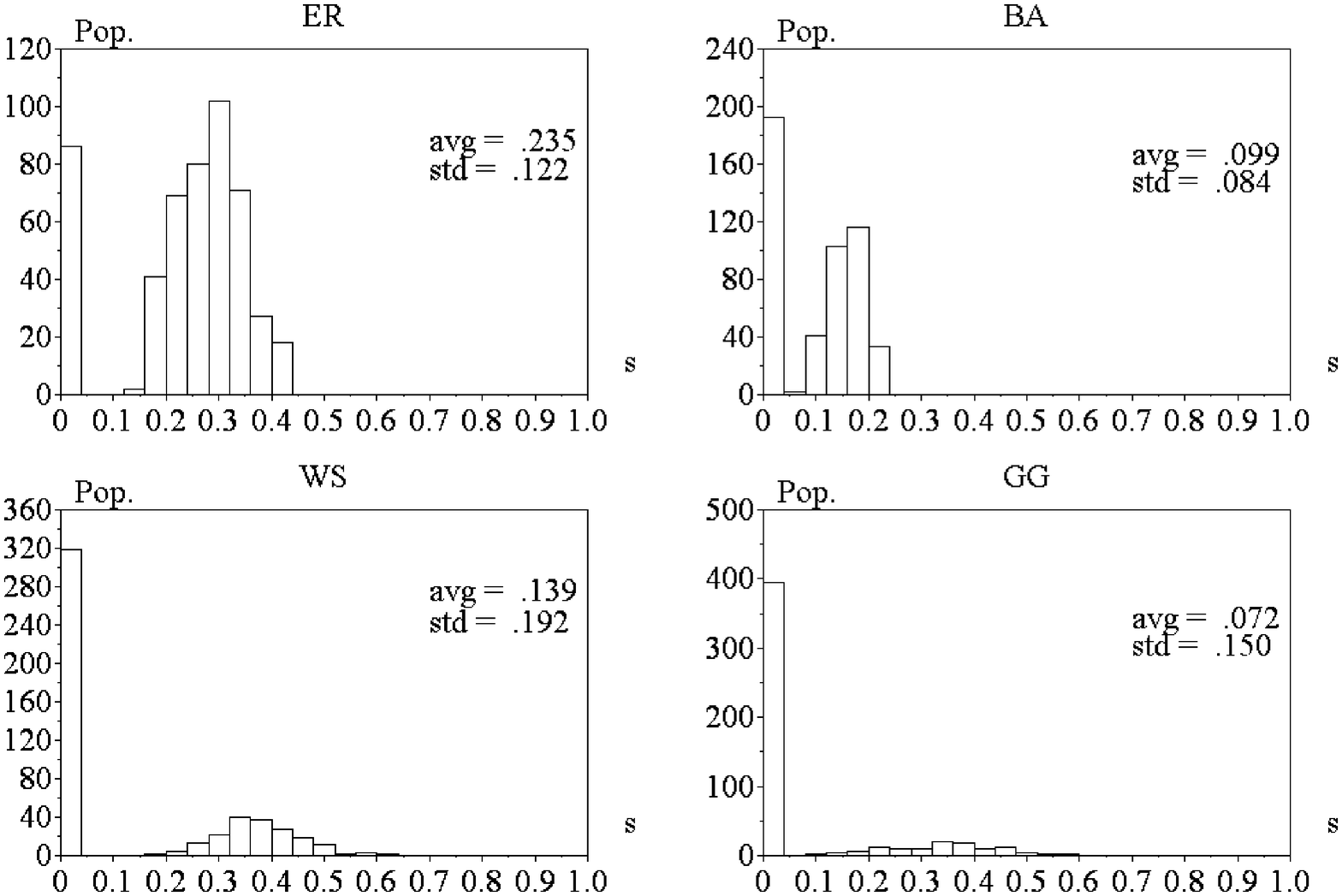} 
  \caption{The population histograms of the separation index $s$
  for the four considered theoretic models of complex networks
  with $N=100$, $M=3$ and $\left< k \right> = 6$.
  A substantial decrease of the attractors separation has
  been implied in all cases.~\label{fig:caso3}}
  \end{center}
\end{figure*}

Now we turn our attention to the influence of the number of prototype
patterns to be represented.  The same configuration as in
Figure~\ref{fig:caso1} is considered, except that $M=10$.  The
obtained results are shown in Figure~\ref{fig:caso3}.  It is clear
that increase by more than threefold of the number of prototype
patterns implied not only a substantially more cases such that $s=0$,
but also the distributions to be strongly left-shifted in comparison
with the respective cases in Figure~\ref{fig:caso1}.  This effect
holds for all the four considered network models. The main reason
behind such a substantial decrease of performance is that simply there
was not much space left, in the average, between the prototype
attractors.

\begin{figure*}[htb]
  \begin{center} 
  \includegraphics[scale=0.55,angle=0]{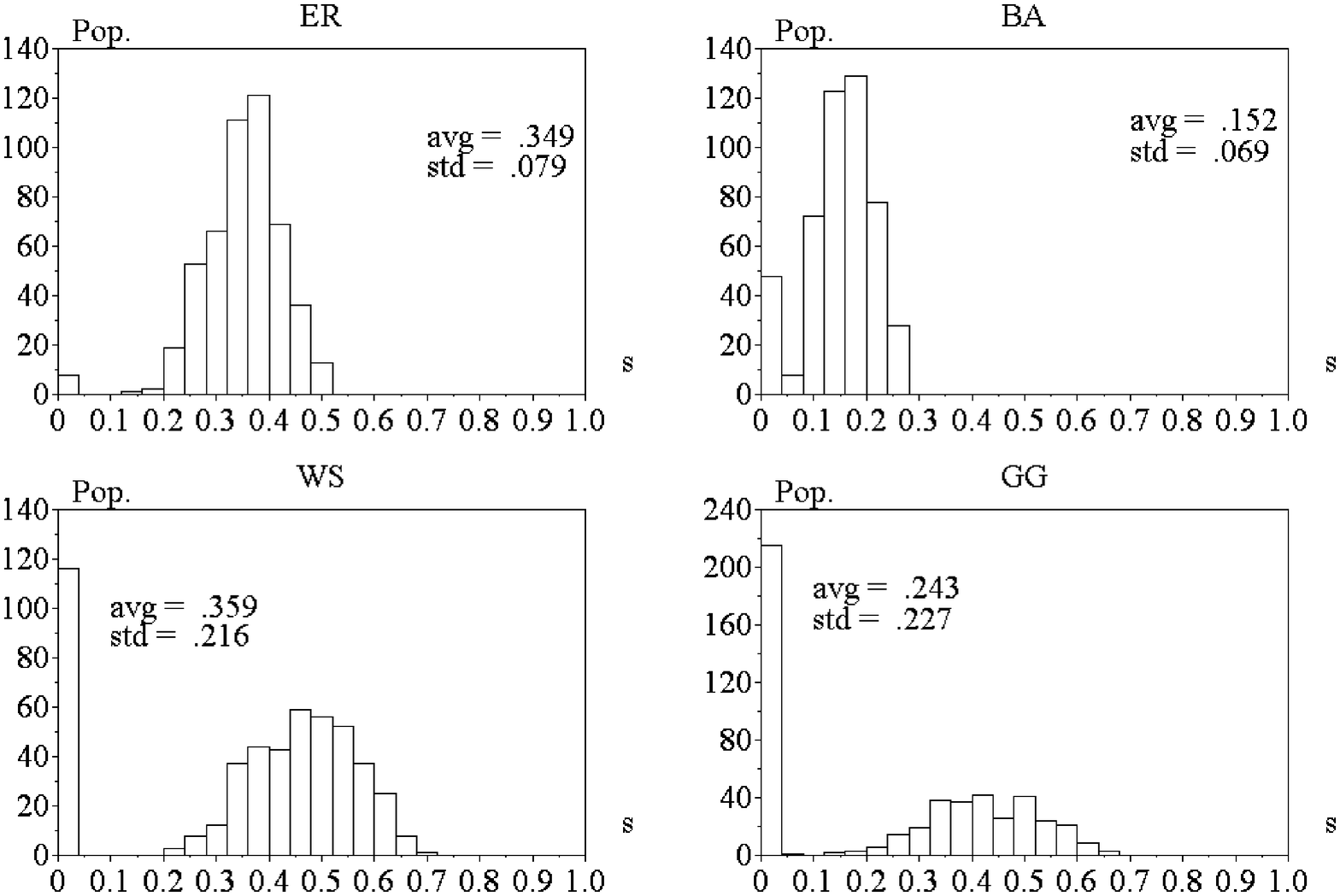} 
  \caption{The population histograms of the separation index $s$
  for the four considered theoretic models of complex networks
  with $N=100$, $M=10$ and $\left< k \right> = 6$.
  A substantial decrease of the attractors separation has
  been implied in all cases.~\label{fig:caso4}}
  \end{center}
\end{figure*}

Because the last configuration exhibited an accentuated loss of
separability because of lack of space in the network, it is
interesting to reconsider the effect of increasing $N$ for this
situation.  The results obtained for the same previous configuration,
but now with $N=200$, are shown in Figure~\ref{fig:caso4}.  The
separation index increased substantially for all the four network
models.  It could be expected that such improvements tend to decrease
for still larger $N$, until reaching a regime where little improvement
is observed.  In such a state, the prototype nodes would be
sufficiently far away one another so that their separation no longer
depends on $N$.  Additional investigation about the change of the
average and standard deviation of $s$ are to be found in
Section~\ref{sec:effs}.

\begin{figure*}[htb]
  \begin{center} 
  \includegraphics[scale=0.55,angle=0]{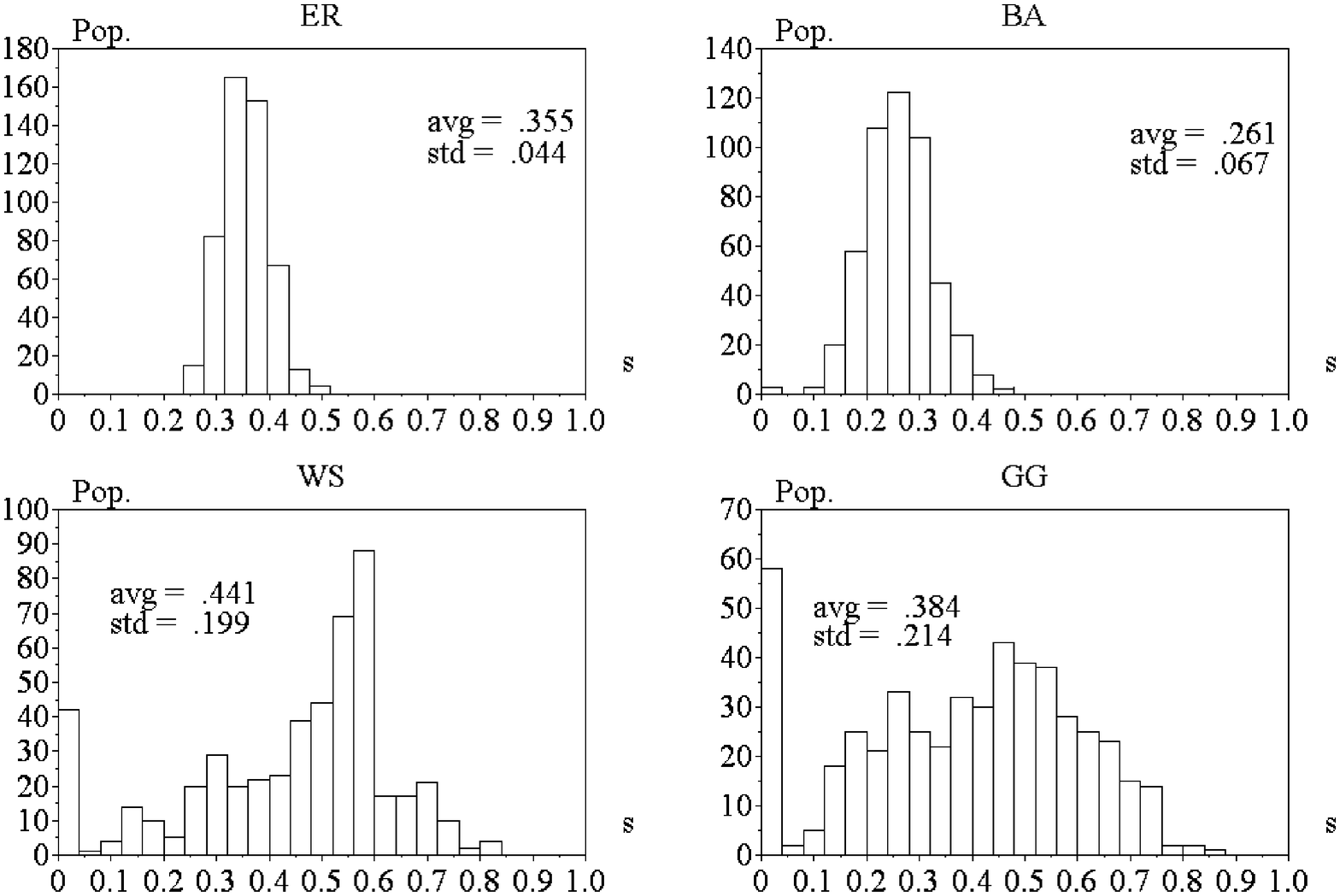} 
  \caption{The population histograms of the separation index $s$
  for the four considered theoretic models of complex networks
  with $N=100$, $M=10$ and $\left< k \right> = 20$.  The effect
  of increasing the average degree was to substantially 
  decrease the attractors separation in all cases.~\label{fig:caso5}}
  \end{center}
\end{figure*}

Finally, we check for the possible effect of the average node degree
on the separation index.  Again, the same configuration as the network
in Figure~\ref{fig:caso1} is adopted, but now with $\left< k
\right> = 20$ (i.e. $m=10$) instead of $\left< k \right> = 6$.  
Figure~\ref{fig:caso5} depicts the respectively obtained results,
which indicate a clear reduction of the attractors separability.  Such
an effect is possibly a consequence of the fact that once the network
become too intensely connected, the shortest path between the
prototype nodes will be reduced and the separability undermined.

\subsection{Finite Size Effects}  \label{sec:effs}

\begin{figure*}[htb]
  \begin{center} 
  \includegraphics[scale=0.55,angle=0]{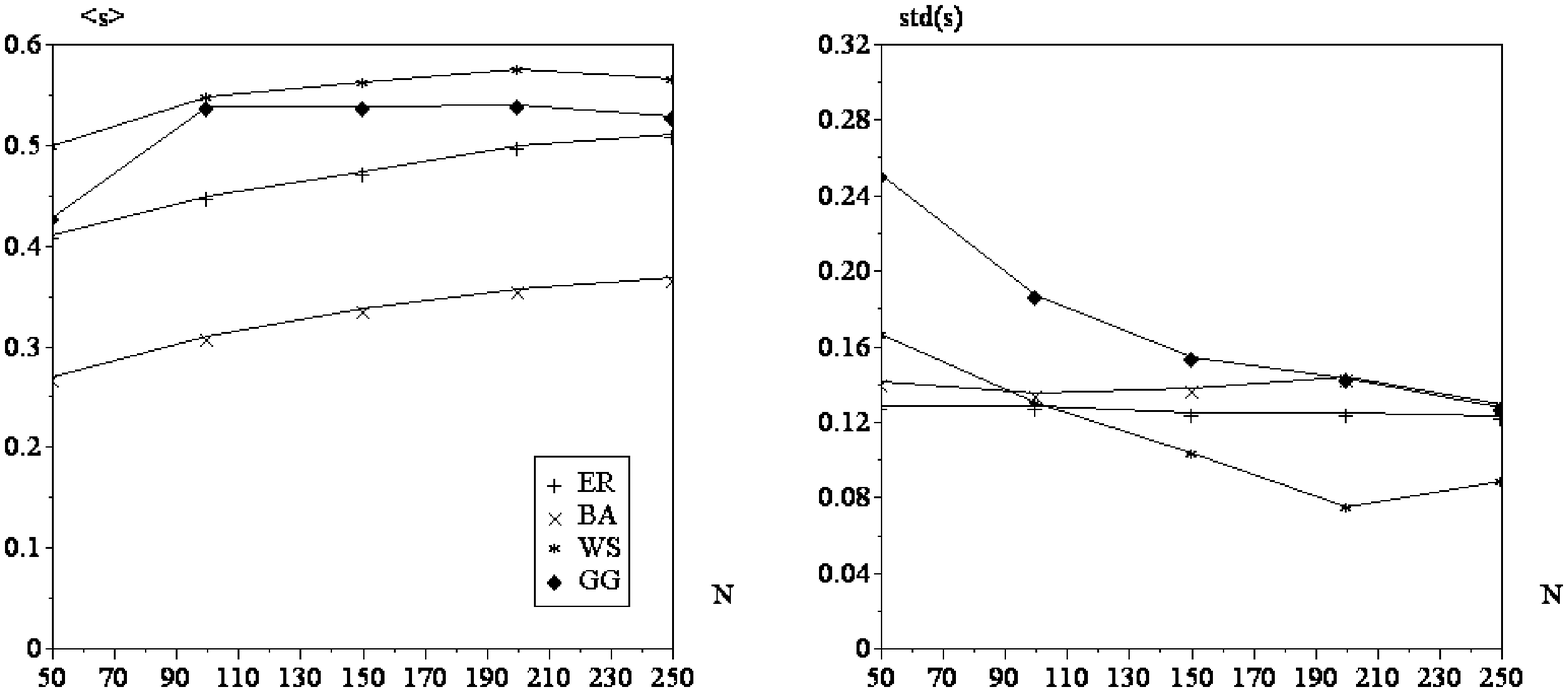} \\
  (a) \hspace{4cm} (b) \\
  \caption{The average and standard deviation of the neighborhood
  separation index ($\left< k \right>$ and $\sigma(s)$, respectively)
  as a function of $N$ for the four models of networks considering
  $m=3$ and $M=3$.  While the averages increase and then tend to 
  a plateau, the standard deviations decrease and also stabilize.
  ~\label{fig:finite}}
  \end{center}
\end{figure*}

The simulations discussed in the previous section seem to have
indicated that increases of the value of $N$ would tend to improve the
attractors separation until reaching a regime where so much space is
available that the patterns no longer feels the finite size of the
network.  In order to obtain further insights about this effect, the
configuration involving $m=3$ and $M=3$ was simulated for $N = 50,
100, 150, 200, 250$ and the results are given in
Figure~\ref{fig:finite}.  This figure shows the average (a) and
standard deviation (b) of the neighborhood separation index in terms
of $N$.  It is clear from these results that the increase of $N$ does
enhance the attractors separation until reaching a plateau, where a
possible 'unsaturation' occurs, indicating that the finite size
effects are over for the specific parameters $m$ and $M$.  At the same
time, $\sigma(s)$ decreases, also reaching a relatively low plateau.
Similar results could be expected for other configurations.  In the
case of a larger value of $M$, it is expected that $\left< k
\right>$ will increase more steeply along the smaller values of $N$,
reaching a similar plateau at larger values.  Therefore, as the
'unsaturation' effect was further corroborated by the additional
analysis, such a behavior can be considered as a guideline for
choosing a proper value of $N$ given $M$ and $m$, for instance by
choosing $N$ where $\left< k
\right>$ reaches a fixed proportion of its plateau value.

\subsection{Correlation Analysis}

In addition to studying the effects of the involved topologic
measurements on the attractors separation, it is also particularly
useful to try to identify in which ways the separation index is
related to them.  This may allow the prediction of the separability
without performing dynamic simulation, i.e. by considering only the
topologic measurements.  Table~\ref{tab:pears} show the Pearson
correlation coefficients obtained considering all pair os topologic
measurement (i.e. $\left<cc\right>$, $\left<sp\right>$, $\sigma(sp)$,
$Vi$, $\left<mi\right>$, $\sigma(mi)$ and $diam$) and the two
separation indices ($s_{ind}$ and $s$).  

As previously verified in~\cite{Costa:survey}, the pattern of
correlations resulted not similar for each model of network. In
addition, the obtained results indicate several high absolute
correlation values.  In the case of $ER$ networks, the most intense
negative correlation (-0.79) was obtained between $\left<sp\right>$
and $\left<mi\right>$, which was indeed expected as longer shortest
path lengths tend to reduce the matching index.  While similar high
correlation (-0.78) was observed also for $BA$, this specific
correlation was relatively smaller for the $WS$ and $GG$ cases (-0.50
and -0.42, respectively).  This fact was reflected in the intense
negative correlation (-0.77) between $s$ and $\left<mi\right>$ and
positive correlation (0.63) with $\left<sp\right>$ for the $ER$ model.
A similar effect can be observed for the $BA$, $WS$ and $GG$ networks.
Observe that $\left<mi\right>$ is highly correlated (0.72 for ER) with
$\sigma(mi)$, suggesting a dependence between the standard deviation
and average of this measurement. The index quantifying the uniformity
of the Voronoi tessellation was found to be negatively correlated with
$\left<mi\right>$ and $\sigma(mi)$, as expected, because higher $mi$
values are favored by more irregular Voronoi areas.  Interestingly,
low correlation values were observed between $s$ and the network
diameter and clustering coefficients in all cases.  It is also
interesting to note that rather different patterns of correlations
were obtained between $s_{ind}$ and $s$ and the topologic measurements
for all network models.

\begin{table*}[htb]
\centering
\vspace{1cm}
{\bf ER} \hspace{0.5cm}
\begin{tabular}{||c||c|c|c|c|c|c|c|c|c||}
 \hline  \hline
 &  $s_{ind}$  &  $s$ & $\left<cc\right>$ & $\left<sp\right>$ & 
   $\sigma(sp)$ & $Vi$ & $\left<mi\right>$ & 
   $\sigma(mi)$ & $ diam $   \\   \hline
 $s_{ind}$         & 1   &     &     &     &     &     &     & &  \\  \hline
 $s$               &-0.11& 1   &     &     &     &     &     & &  \\  \hline
 $\left<cc\right>$ &-0.06&-0.11& 1   &     &     &     &     & &  \\  \hline
 $\left<sp\right>$ &-0.51& 0.63&-0.11& 1   &     &     &     & &  \\  \hline
 $\sigma(sp)$      & 0.31&-0.38& 0.07&-0.11& 1   &     &     & &  \\  \hline
 $Vi$              &-0.17& 0.40&-0.03& 0.19&-0.19& 1   &     & &  \\  \hline
 $\left<mi\right>$  & 0.51&-0.77& 0.08&-0.79& 0.49&-0.44& 1 & &  \\ \hline
 $\sigma(m)i$       & 0.37&-0.52& 0.09&-0.51& 0.44&-0.52& 0.72& 1 &  \\  \hline
 $diam$            & 0.02& 0.00&-0.06& 0.03&-0.01& 0.10&-0.01&-0.05& 1 \\ \hline  \hline
\end{tabular} \hspace{0.5cm} (a) \vspace{0.3cm} \\
{\bf BA} \hspace{0.5cm}
\begin{tabular}{||c||c|c|c|c|c|c|c|c|c||}  
 \hline  \hline
 &  $s_{ind}$  &  $s$ & $\left<cc\right>$ & $\left<sp\right>$ & 
   $\sigma(sp)$ & $Vi$ & $\left<mi\right>$ & 
   $\sigma(mi)$ & $ diam $   \\   \hline
 $s_{ind}$         & 1   &     &     &     &     &     &     & &  \\  \hline
 $s$               &-0.02& 1   &     &     &     &     &     & &  \\  \hline
 $\left<cc\right>$ &-0.08&-0.26& 1   &     &     &     &     & &  \\  \hline
 $\left<sp\right>$ &-0.53& 0.62&-0.13& 1   &     &     &     & &  \\  \hline
 $\sigma(sp)$      & 0.25&-0.10&-0.16& 0.01& 1   &     &     & &  \\  \hline
 $Vi$              &-0.10& 0.81&-0.26& 0.54&-0.12& 1   &     & &  \\  \hline
 $\left<mi\right>$  & 0.44&-0.69& 0.13&-0.78& 0.40&-0.65& 1 & &  \\ \hline
 $\sigma(mi)$       & 0.47&-0.35& 0.06&-0.58& 0.28&-0.46& 0.61& 1 &  \\  \hline
 $diam$            & 0.03& 0.09&-0.24& 0.08&-0.08& 0.07&-0.06&-0.03& 1 \\ \hline  \hline
\end{tabular} \hspace{0.5cm} (b) \vspace{0.3cm} \\
{\bf WS} \hspace{0.5cm}
\begin{tabular}{||c||c|c|c|c|c|c|c|c|c||}   
 \hline  \hline
 &  $s_{ind}$  &  $s$ & $\left<cc\right>$ & $\left<sp\right>$ & 
   $\sigma(sp)$ & $Vi$ & $\left<mi\right>$ & 
   $\sigma(mi)$ & $ diam $   \\   \hline
 $s_{ind}$         & 1   &     &     &     &     &     &     & &  \\  \hline
 $s$               &-0.58& 1   &     &     &     &     &     & &  \\  \hline
 $\left<cc\right>$ &-0.04& 0.09& 1   &     &     &     &     & &  \\  \hline
 $\left<sp\right>$ &-0.46& 0.54& 0.10& 1   &     &     &     & &  \\  \hline
 $\sigma(sp)$      & 0.43&-0.16& 0.06& 0.31& 1   &     &     & &  \\  \hline
 $Vi$              &-0.55& 0.63& 0.10& 0.53&-0.09& 1   &     & &  \\  \hline
 $\left<mi\right>$  & 0.89&-0.80&-0.04&-0.50& 0.36&-0.69& 1 & &  \\ \hline
 $\sigma(mi)$       & 0.42&-0.54&-0.05&-0.37& 0.13&-0.37& 0.55& 1 &  \\  \hline
 $diam$            &-0.01& 0.02& 0.29& 0.18& 0.16& 0.07&-0.02&-0.02& 1 \\ \hline  \hline
\end{tabular} \hspace{0.5cm} (c) \vspace{0.3cm} \\
{\bf GG} \hspace{0.5cm}
\begin{tabular}{||c||c|c|c|c|c|c|c|c|c||}  
 \hline  \hline
 &  $s_{ind}$  &  $s$ & $\left<cc\right>$ & $\left<sp\right>$ & 
   $\sigma(sp)$ & $Vi$ & $\left<mi\right>$ & 
   $\sigma(mi)$ & $ diam $   \\   \hline
 $s_{ind}$         & 1   &     &     &     &     &     &     & &  \\  \hline
 $s$               &-0.16& 1   &     &     &     &     &     & &  \\  \hline
 $\left<cc\right>$ &-0.06& 0.04& 1   &     &     &     &     & &  \\  \hline
 $\left<sp\right>$ &-0.04& 0.42& 0.13& 1   &     &     &     & &  \\  \hline
 $\sigma(sp)$      & 0.16&-0.10& 0.11& 0.57& 1   &     &     & &  \\  \hline
 $Vi$              &-0.38& 0.71& 0.08& 0.40&-0.09& 1   &     & &  \\  \hline
 $\left<mi\right>$  & 0.40&-0.87&-0.06&-0.42& 0.16&-0.71& 1 & &  \\ \hline
 $\sigma(mi)$       & 0.33&-0.72&-0.06&-0.31& 0.20&-0.58& 0.82& 1 &  \\  \hline
 $diam$            &-0.07& 0.04& 0.19& 0.31& 0.30& 0.12&-0.10&-0.05& 1 \\ \hline  \hline
\end{tabular} \hspace{0.5cm} (d) \vspace{0.3cm}
\caption{Pearson correlation coefficients obtained for the
ER (a), BA (b), WS (c) and GG (d) networks with $N=100$, $m=3$ and
$M=3$ considering the individual ($s_{ind}$) and neighborhood
($s_{ngh}$) separability indices and seven measurements of the
topology of the network (i.e. average clustering coefficient
($\left<cc\right>$), average shortest path between prototype nodes
($\left<sh_path\right>$), standard deviation of the shortest paths
between the prototype nodes ($\sigma(sh_path)$)), average Voronoi
index ($Vi$), average matching index between prototype
nodes ($\left<m\right>$), standard deviation of the matching index
between prototype nodes ($\sigma(mi)$) and diameter of the networks
(diam)}.\label{tab:pears}
\end{table*}

\subsection{Structural Equation Modeling (SEM) / Path Analysis}  
\label{sec:res_path}

Although the correlation analysis described in the previous section
can provide interesting information about the pairwise relationship
between the separation indices and the topologic measurements, such
results are limited because they do not reflect the general
relationship between all the topologic measurements and the separation
indices.  In addition, some of the high observed absolute
correlation values can be a consequence of
spurious~\cite{Raykov:2000,Kline:2005} effects between the variables.
In order to gather additional insights about the how the dynamic
parameters (i.e. the separation indices) are influenced (and even to a
large extent defined) by the topology of the network, while
considering all measurement co-variations, we performed path analysis
considering the structural dependence between measurements as
expressed in Figure~\ref{fig:diagram}.  It is expected that, by
considering all dispersions, the path analysis can provide a more
objective and filtered indication of the influences of the topologic
measurements on the attractors separation.

The considered data were respective to ER, BA, WS and GG networks with
$N=100$, $M=3$ and $m=3$. After estimation of the covariance of the
measurements, coding and execution in the LISREL environment, the
regression coefficients and residues ($E1$ and $E2$), as well as the
relationship between these residuals ($C_{12}$) were obtained.  The
results are given in Table~\ref{tab:pathan}.  

A series of interesting insights have been derived from these results.
As with the correlations, the dependencies between the topology and
separation are strictly specific to each network model, a dependency
which may also change for other configurations with different values
of the parameters $M$ and $m$.  In the case of the ER networks, the
standard deviation of the matching index ($\sigma(mi)$) resulted
particularly influent (negative influence = -0.96) on the $s$ index.
At the same time, the average matching index ($\left<k\right>$) was
found to have a strong influence on $s$.  Except for relatively
smaller negative influence of the $\left< cc
\right>$ on $s$, no particularly strong influences are observed with
respect to the other topologic measurements.  Such a strong influence
of the matching index can be understood because non-zero matching
index are obtained only in extreme cases, where the prototype nodes
are too close.  Therefore, nonzero matching indices are a strong and
secure indication of poorly separated attractors.  For this reason,
the matching index dominated the path analysis and implied smaller
influences for most other measurements, including the shortest path.
This effect can also be observed for the BA and WS cases.  However, it
is interesting to note the weak influence of $\sigma(mi)$ on $s$ in
the case of GG networks.  For the BA case, the strongest influence on
$s$ was identified for the Voronoi index $Vi$, which is compatible
with the fact that higher uniformity of the Voronoi tessellation by
the prototype nodes tends to promote better separation between those
nodes.  Interestingly, a particularly strong influence of $Vi$ on the
$s$ has been observed only for the BA model.  This effect can be
related to the fact that the BA provides the poorest general
separation between attractors as a possible consequence of the
generalized connectivity implemented by the hubs.  In such cases,
where a larger number of non-zero matching indices are therefore
obtained, the Voronoi separation may become more relevant as a
predictor of the separability.  In the case of WS models, strong
influence (positive = 0.65) was obtained for the $\left<cc\right>$.
This effect is particularly interesting because it could be expected
that, by promoting smaller shortest paths, this measurement would be
inversely related to the separability, which is indeed the case for
the respective ER and BA path analysis results.  Indeed, a very weak
dependency with this variable had been revealed by the Pearson
correlations.  However, the positive influence of $\left<cc\right>$ on
$s$ can be a consequence of the fact that GG networks with higher
average clustering coefficients will tend to have more intense local
connectivity which could allow the activity to diffuse more uniformly
and to concentrate effectively around the prototype nodes.  Such an
effect would be more definite in the WS case because of the higher
uniformity of local connections implied by its ring structure.  The
influences of the topologic features on the $s$ obtained for the GG
cases are dominated by $\left<mi\right>$ as discussed above.

\begin{table*}[htb]
\centering
\vspace{1cm}
\begin{tabular}{|c|c||c|c|c|c|c|c|c|c|c|}
  \hline  
    & &  $\left<cc\right>$ & $\left<sp\right>$ & 
    $\sigma(sp)$ & $ Vi$ & $\left<mi\right>$ & 
    $\sigma(mi)$ & $ diam $ &  $Res.$ &  $C$ \\   \hline  \hline
    ER  & 
    $s_{ind}$   & $\gamma_{13}=-0.01$ & $\gamma_{14}=-0.02$ & 
                $\gamma_{15}=0.01$    & $\gamma_{16}=-0.01$ 
                & $\gamma_{17}=0.01$ & $\gamma_{18}=0.01$ & 
                $\gamma_{19}=0.00$ & $E1=0.00$ & $C_{1,2}=0.01 $ \\ 
    & $s$       & $\gamma_{23}=-0.43$ & $\gamma_{24}=0.04$ & 
                $\gamma_{25}=-0.03$ & $\gamma_{26}=0.25$ & 
                $\gamma_{27}=-0.96$ & $\gamma_{28}=1.13$ & 
                $\gamma_{29}=-0.01$ & $E2=0.00$ & \\  \hline \hline
    BA  & 
    $s_{ind}$   & $\gamma_{13}=-0.02$ & $\gamma_{14}=-0.02$ & 
                $\gamma_{15}=0.01$    & $\gamma_{16}=0.03$ 
                & $\gamma_{17}=-0.02$ & $\gamma_{18}=0.06$ & 
                $\gamma_{19}=0.00$ & $E1=0.00$ & $C_{12}=0.00$ \\ 
    & $s$       & $\gamma_{23}=-0.21$ & $\gamma_{24}=0.06$ & 
                $\gamma_{25}=0.01$ & $\gamma_{26}=0.75$ & 
                $\gamma_{27}=-0.39$ & $\gamma_{28}=0.73$ & 
                $\gamma_{29}=0.00$ & $E2=0.00$ & \\  \hline \hline
    WS  & 
    $s_{ind}$   & $\gamma_{13}=-0.05$ & $\gamma_{14}=0.00$ & 
                $\gamma_{15}=0.01$    & $\gamma_{16}=0.03$ 
                & $\gamma_{17}=0.17$ & $\gamma_{18}=-0.69$ & 
                $\gamma_{19}=0.00$ & $E1=0.00$ & $C_{12}=0.00$\\ 
    & $s$       & $\gamma_{23}=0.65$ & $\gamma_{24}=0.01$ & 
                $\gamma_{25}=0.01$ & $\gamma_{26}=0.08$ & 
                $\gamma_{27}=-0.55$ & $\gamma_{28}=-3.54$ & 
                $\gamma_{29}=-0.16$ & $E2=0.00$ & \\  \hline  \hline
    GG  & 
    $s_{ind}$   & $\gamma_{13}=-0.03$ & $\gamma_{14}=0.00$ & 
                $\gamma_{15}=0.00$    & $\gamma_{16}=-0.02$ 
                & $\gamma_{17}=0.05$ & $\gamma_{18}=-0.01$ & 
                $\gamma_{19}=0.00$ & $E1=0.00$ & $C_{12}=0.01$ \\ 
    & $s$       & $\gamma_{23}=-0.10$ & $\gamma_{24}=0.00$ & 
                $\gamma_{25}=0.00$ & $\gamma_{26}=0.17$ & 
                $\gamma_{27}=-0.87$ & $\gamma_{28}=-0.08$ & 
                $\gamma_{29}=-0.01$ & $E2=0.00$ & \\  \hline 
\end{tabular}
\caption{The influences of each considered topologic measurement
on the attractors separation indices as revealed by path
analysis considering $N=100$,$M=3$ and $m=3$.}\label{tab:pathan}
\end{table*}

While the previous path analysis has revealed a series of insights
about the influence of the network topology on the dynamic separation
of its attraction basins, it was strongly biased by the critical
influence of the matching index.  In order to try to get further
insights on the structure/dynamics relationship for the four
considered models, we repeated the path analysis while not including
$\left<mi\right>$ and $\sigma(mi)$.  The results are summarized in
Table~\ref{tab:pathan2}.  Interestingly, except for relatively small
increases with respect to $Vi$ and $\left<cc\right>$, the obtained
influences remained similar, with small effects observed for the
shortest path measurements.  This can be understood as a confirmation
that the separation between the attractors is critically defined by
the overlap between the hierarchical neighbors of the prototype
nodes, especially their immediate neighbors (implying higher matching
indices).  In the cases where the matching index is zero, the lack of
congruence between the original attractors and the distribution of
diffusive activity should be largely explained by the higher
influences observed in Table~\ref{tab:pathan2}, which involve the
average clustering coefficient of the network and the Voronoi index
for the prototype nodes.  Indeed, in the cases where the prototype
nodes are not topologically too close one another, the lack of
agreement between the activity distribution and the original
attractors (as in Figure~\ref{fig:ex}b and c) will depend particularly
on the properties of the local connectivity as expressed by the
clustering coefficient and the node degree.

\begin{table*}[htb]
\centering
\vspace{1cm}
\begin{tabular}{|c|c||c|c|c|c|c|c|c|}
  \hline  
    & &  $\left<cc\right>$ & $\left<sp\right>$ & 
    $\sigma(sp)$ & $Vi$ & $ diam $ &  
    $Res.$ &  $C$ \\   \hline  \hline
    ER  & 
    $s_{ind}$   & $\gamma_{13}=-0.01$ & $\gamma_{14}=-0.02$ & 
                $\gamma_{15}=0.01$    & $\gamma_{16}=-0.01$ 
                & $\gamma_{17}=0.00$ & $E1=0.00$ & $C_{12}=0.00$  \\ 
    & $s$       & $\gamma_{23}=-0.26$ & $\gamma_{24}=0.14$ & 
                $\gamma_{25}=-0.09$ & $\gamma_{26}=0.43$ & 
                $\gamma_{27}=-0.01$ & $E2=0.00$ & \\  \hline \hline
    BA  & 
    $s_{ind}$   & $\gamma_{13}=-0.02$ & $\gamma_{14}=-0.02$ & 
                $\gamma_{15}=0.01$    & $\gamma_{16}=0.03$ 
                & $\gamma_{17}=0.00$ & $E1=0.00$ & $C_{12}=0.00$ \\ 
    & $s$       & $\gamma_{23}=-0.22$ & $\gamma_{24}=0.08$ & 
                $\gamma_{25}=-0.01$ & $\gamma_{26}=0.79$ & 
                $\gamma_{27}=-0.00$ & $E2=0.00$ & \\  \hline \hline
    WS  & 
    $s_{ind}$   & $\gamma_{13}=0.03$ & $\gamma_{14}=-0.01$ & 
                $\gamma_{15}=0.01$    & $\gamma_{16}=-0.05$ 
                & $\gamma_{17}=0.00$ & $E1=0.00$ & $C_{12}=0.00$\\ 
    & $s$       & $\gamma_{23}=0.43$ & $\gamma_{24}=0.03$ & 
                $\gamma_{25}=-0.03$ & $\gamma_{26}=0.41$ & 
                $\gamma_{27}=-0.01$ & $E2=0.00$ & \\  \hline  \hline
    GG  & 
    $s_{ind}$   & $\gamma_{13}=-0.01$ & $\gamma_{14}=0.00$ & 
                $\gamma_{15}=0.0$    & $\gamma_{16}=-0.04$ 
                & $\gamma_{17}=0.00$ & $E1=0.00$ & $C_{12}=0.01$ \\ 
    & $s$       & $\gamma_{23}=-0.21$ & $\gamma_{24}=0.04$ & 
                $\gamma_{25}=-0.03$ & $\gamma_{26}=0.51$ & 
                $\gamma_{27}=-0.01$ & $E2=0.00$ & \\  \hline 
\end{tabular}
\caption{The influences of each considered topologic measurement
on the attractors separation indices as revealed by path
analysis considering $N=100$,$M=3$ and $m=3$.}\label{tab:pathan2}
\end{table*}

\section{Conclusions}

The investigation about the relationship between the structure and
function of networks (e.g.~\cite{Newman03, Boccaletti05,
Costa_Salou:2005}) represents one of the most interesting perspectives
for obtaining insights about complex dynamic systems.  As briefly
reviewed in this article, several works have addressed the problem of
how the structure of the connectivity may affect and largely define
the properties of dynamic systems.  One particularly important aspect
which has received relatively lesser attention concerns the
separability between different grandmother attractors, each
representing a prototype pattern or state.  This issue is critically
relevant because it has great impact on the capacity of the network
for proper representation of patterns, the level of
redundancy/robustness of such representations, the degree of
generalization for recognition of not previously trained prototypes,
as well as the effectiveness during retrieval and activation of such
prototypes.  While other types of coding can be used in dynamic
system, grandmother representation stands out as particularly
important because it seems to be the way a great part of the primates
cortex is organized.  

The current work has reported an approach to the characterization of
the separability between prototype patterns which incorporates a
number of special features.  First, we have considered four
representative theoretic models of complex networks -- namely the
random networks of Erd\H{o}-R\'enyi, the scale free model of
Barab\'asi-Albert, the small-world networks of Watts-Strogatz, as well
as a simple, non-small-world topographic model.  Second, by using a
generic diffusive process, followed by the calculation of the
respective derivative network in order to obtain the attraction
basins, we obtained a methodology which completely avoids the
intricacies and specifities implied by each type of dynamic system.
While it remains to be verified how good accurate and general such an
approximation is, it does allow the definition of smooth attraction
basins around each prototype node in a way which is remindful of many
important dynamical systems such as Kohonen's self-organizing maps and
the primates cortex.  Once the attraction basins are so defined, a
simple diffusive scheme emanating from each network node is employed
in order to obtain the general activation of the network.  Although
ideally such a procedure should activate equally only the original
prototype nodes, the structured connectivity of the networks will act
so as to produce non-uniform activation, where just a fraction of the
overall activation coincides with the prototype nodes.  The
disagreement between the original prototypes and the induced activity
has been quantified in terms of two separation indices, namely the
individual and neighborhood indices.  Because the former consider only
the agreement and balance between the induced activation and the
original prototypes, it resulted to be too small and therefore with
low resolution for quantifying the attractors separation.  By
considering also the immediate neighborhood of the prototype nodes,
the second index allowed a more informative indication of the
attractors separability.

The specific topologic features of each of the four considered
theoretic network models can have different effects for the separation
of the attraction basins.  Therefore, we performed an investigation of
the effects of the most important parameters of the simulations over
the respective performance.  We verified that the separation tends to
decrease with the number of prototypes and increase with the size of
the network.  At the same time, more intense general connectivity, as
expressed by the average node degree, also tended to undermine the
attractors separation.  Special attention was given to the finite
sizes implied by the parameter $N$.  The obtained results seem to
suggest that the separability tends to increase with $N$ up to a
regime where the finite size of the network is no longer felt
(`unsaturation').  Therefore, by identifying the region where such a
plateau of separability is reached, it is possible to obtain
near-optimum values of $N$ for given $M$ and $m$.  Next, we applied
correlation analysis in orde not only to identify the
inter-relationships between each topologic measurement, but also
between these and the two separation indices.  The presence of such
correlations can not only help to understand the origin of the
attractors separation, but also allow the prediction of such a
property from measurements of the network topology, without the need
of simulations.  Several tendencies were identified through such an
analysis, including the tendency of the index $s$ to be strongly
proportional to the uniformity of the Voronoi areas, as expressed by
the Voronoi index $Vi$.  A positive, though weaker, correlation was
also identified between $s$ and the average shortest path length
$\left<sp\right>$ of the networks.  A strong negative correlation was
identified between $s$ and the matching index $mi$.  Although all such
behaviors are compatible with what could be expected, meaningfully
different correlations differences were observed for each network
model.  Interestingly, the network model allowing the best overall
separation of attractors was found to be the Watts-Strogatz, followed
by the geographic, random and scale-free models.  It is conjectured
here that the superior properties of the WS model stem from its
enhanced uniformity and low randomness of local connectivity as well
as by the fact that, although being a small world model, there are
relatively very few long range connections between any pair of nodes.
Similar properties, though at the expense of a weaker local order and
uniformity of local connections, are characteristic of the geographic
model, which came in second in performance.  As a matter of fact, the
WS and GG models showed similar behavior as far as the histograms of
separation index were concerned.  The same was observed for the ER and
BA cases.  Therefore, from the perspective of attractors separation,
the WS/GG and ER/BA models seem to represent two different classes of
systems, with the latter providing rather poorer separability.
Interestingly, the small world property can not explain such a
partition, as the GG is not a small world model as the other three
cases.  Consequently, it seems that the common denominators in the
pairs WS/GG and ER/BA seems to be more strongly related also to the
uniformity of the local connectivity.  Such an explanation would be
largely in agreement with several of the results of works
investigating the effect of connectivity on memory, as reviewed in
Section~\ref{sec:review}.

In order to try to learn more about the effects of the topological
features of the network on the dynamical property of attractors
separation, path analysis was also applied assuming a simple
structural relationship.  To our best knowledge, this is the first
time such an insightful analysis has been applied for the study of
complex networks.  One of the main advantages of such an approach over
the Pearson correlation coefficients is that here the dispersions of
\emph{all} the considered measurements are taken into account instead
of the pairwise relationships underlying the correlations.
Interestingly, the path analysis yielded influences of the topologic
measurements on the separation indices which were often substantially
different from those suggested by the Pearson correlation
coefficients.  Of particular interest was the near null influence
assigned to the average and standard deviation of the shortest path
length between the prototype nodes.  This is all the most surprising
as it seems intuitively reasonable to discuss much of the effect of
the topology of the network over the attractors separation by
considering such measurements.  More specifically, networks where the
prototype nodes are topologically close one another would tend to have
poorer separation.  However, the performed path analysis substantially
emphasized, for all the four considered models, the importance of the
average matching index on the attractors separation.  On second
thoughts, this is indeed reasonable because the presence of non-zero
matching index is a certain indication of overlap of the attraction
basins.  Even when repeating the path analysis while leaving out the
matching index measurements, the other parameters (especially the
shortest path length) did not result with higher influences.  Other
interesting insights, including influences which were specific to
network models, were also allowed by the path analysis, corroborating
therefore the potential value of such a statistical approach in order
to get insights about the relationship between structure and dynamics
of complex dynamic systems.  However, it is important to keep in mind
that every statistical methodology should be understood as a source of
insights to be further investigated and corroborated rather than
spelling definitive facts.

Despite the relative comprehensiveness of the presently reported
investigation, and perhaps as its consequence, a series of future
developments can be suggested.  First, it would be interesting to
extend the reported investigation to larger network sizes and to
consider additional measurements such as the standard deviation of the
clustering coefficient and the spectral structure of the adjacency and
weight matrix.  Second, the sources of the deviations of the induced
activation from the prototype nodes could be further investigated by
considering simulations involving a single attractor.  In such cases,
all the loss of `separability' would necessarily be a consequence of
the probability leakage from the prototype node into its neighbors.
It would be particularly interesting to verify, through path analysis,
which of the topological measurements of the network would be more
influence on the attractor activation.  It would also be worth
investigating the potential of using the \emph{percolation
transform}~\cite{Costa_percol:2004, Costa_bioinfo:2005} over the
network as the means to explain and predict the attractors
separability.  Other promising possibilities include the extension of
the matching index to take into account hierarchical levels larger
than one, i.e. including also the second and higher neighborhoods.

Several real problems are closely related to the issue of attractor
separation and prototype activation.  One real problem which could be
particularly interesting to be addressed is the phenomenon of
\emph{facilitation} of neurons (represented by nodes).  By
facilitation of a neuron it is meant that that neuron will become more
likely to engage into activity.  For instance, the definition of
temporary priorities in the primates brain could be related to the
reinforcement of one or more attractors, so that they become more
likely to be revised along time (e.g. through a random walk).
Interestingly, as suggested by the current work, the facilitation of a
given node would be highly dependent on its local connectivity.  Such
studies could be eventually extended to higher level mental dynamics,
such as those underlying attention and even pathologies
(e.g.~\cite{Wedemann:2006}). It is also reasonable to expect that the
definition of attractors is a process which co-evolves with the
network topology.  In this sense, it would be interesting to try to
identify growing schemes where the connectivity is affected by the
success of the establishment of prototype attractores, and vice-versa.
Another related investigation would be the identification of topologic
organizations of networks allowing optimal or near-optimal attractors
separation.

\vspace{1cm}
{\bf List of Symbols}

\begin{trivlist}

\item  $\Gamma =$ a graph or complex network;

\item  $N =$ number of nodes in a network;

\item  $K =$ the adjacency matrix of a complex network;

\item  $k(i) =$ degree of a network node $i$;

\item  $m =$ the parameter in the BA model defining its average
               node degree.  This parameter is considered as
               reference for all network models in this work;

\item  $diam =$ the diameter of the network;

\item  $cc(i) =$ clustering coefficient of a network node $i$;

\item  $sp(i,j)$ shortest path between nodes $i$ and $j$ in a complex network;

\item  $mi(i,j) = $ the matching index of nodes $i$ and $j$;

\item  $Vi =$ the Voronoi index of separation between the areas of
              influence of the prototype nodes;

\item $\left< a \right> =$ the arithmetic average of the property $a$;

\item $\left[ a \right] =$ the geometric average of the property $a$;

\item $\sigma(a) =$ the standard deviation of the random variable $a$;

\item $M =$ the number of prototype patterns to be represented in a network;

\item $\vec{\alpha} =$ the activations in a network after a long 
      random walk;

\item $s_{ind} =$ the separability index between prototype nodes in a 
      network at the level of individual nodes;

\item $s =$ the separability index between prototype nodes in a network
            considering also the immediate neighborhood of such nodes;

\end{trivlist}

\begin{acknowledgments}
Luciano da F. Costa thanks CNPq (308231/03-1) and FAPESP (05/00587-5)
for sponsorship.
\end{acknowledgments}

\bibliography{mem}

\end{document}